\RequirePackage{ifpdf}
\ifpdf 
\documentclass[pdftex]{sigma}
\else
\documentclass{sigma}
\fi

\font\fr=eufm10 scaled \magstep 1 
\def\derpar#1#2{\frac{\partial{#1}}{\partial{#2}}}

\def\moment#1#2#3{{#1}_{#2}, \ldots, {#1}_{#3}}

\def\vf{\mbox{\fr X}}
\def\df{{\mit\Omega}}
\def\Lag{{\cal L}}
\def\lag{\pounds}

\def\d{{\rm d}}

\def\Real{\mathbb{R}}

\def\inn{\mathop{i}\nolimits}
\def\Tan{{\rm T}}

\def\Lie{\mathop{\rm L}\nolimits}
\def\ls{(J^1\pi,\Omega_\Lag)}
\def\hsjpi{(J^1\pi^*,\Omega_h)}

\def\hsjpio{({\cal P},\Omega_h^0)}

\def\Cinfty{{\rm C}^\infty}

\begin{document}

\allowdisplaybreaks

\renewcommand{\PaperNumber}{100}

\FirstPageHeading

\ShortArticleName{Multisymplectic Lagrangian and Hamiltonian Formalisms of Classical Field Theories}

\ArticleName{Multisymplectic Lagrangian and Hamiltonian\\ Formalisms of Classical Field Theories}

\Author{Narciso ROM\'AN-ROY}

\AuthorNameForHeading{N. Rom\'an-Roy}

\Address{Dept. Matem\'atica Aplicada IV,
  Edif\/icio C-3, Campus Norte UPC,\\
  C/ Jordi Girona 1, E-08034 Barcelona, Spain}

\Email{\href{mailto:nrr@ma4.upc.edu}{nrr@ma4.upc.edu}}
\URLaddress{\url{http://www-ma4.upc.edu/~nrr/}}

\ArticleDates{Received July 02, 2009, in f\/inal form October 30, 2009;  Published online November 06, 2009}

\Abstract{This review paper is devoted to presenting
the  standard  {\it multisymplectic formulation}
for describing geometrically classical f\/ield theories,
both the  regular and singular cases.
First, the main features of the Lagrangian formalism
are revisited and, second,
the Hamiltonian formalism is constructed using {\it Hamiltonian sections}.
In both cases, the va\-riational principles leading to
the Euler--Lagrange and the Hamilton--De Donder--Weyl equations,
respectively, are stated,
and these f\/ield equations are given in dif\/ferent but equivalent
geometrical ways in each formalism.
Finally, both are unif\/ied in a new formulation
(which has been developed in the last years), following
the original ideas of Rusk and Skinner for mechanical systems.}


\Classification{70S05; 55R10; 53C80}

\vspace{-2mm}

\section{Introduction}

In recent years much work has been done
with the aim of establishing the suitable geometrical structures
for describing classical f\/ield theories.

There are dif\/ferent kinds of geometrical models
for making a covariant description of classical f\/ield theories
described by f\/irst-order Lagrangians.
For instance, we have the so-called {\it $k$-symplectic formalism}
which uses the {\it $k$-symplectic forms}
introduced by Awane \cite{Aw-92,aw2,aw3},
and which coincides with the {\it polysymplectic formalism}
described by  G\"unther \cite{Gu-87} (see also \cite{fam}).
A~natural extension of this is the
{\it $k$-cosymplectic formalism}, which is the generalization
to f\/ield theories of the cosymplectic description of
non-autonomous mechanical systems \cite{mod1,LMS-99}.
Furthermore, there are the polysymplectic formalisms
developed by Sardanashvily~et al.~\cite{GMS-97,Sd-95} and Kanatchikov \cite{Ka-98},
which are based on the use of vector-valued forms on
f\/iber bundles, and which are dif\/ferent descriptions of classical
f\/ield theories than the polysymplectic one proposed by G\"unther.
In addition, soldering forms on linear
frame bundles are also polysymplectic forms, and their study and
applications to f\/ield theory constitute the $k$-symplectic
geometry developed by Norris \cite{McN,No2,No5}.
There also exists the formalism based on using {\it Lepagean forms},
used for describing certain kinds of equivalent Lagrangian models
with non-equivalent Hamiltonian descriptions
 \cite{Kr-87,Krva-2002,KS-01a,KS-01b}.
Finally, a new geometrical framework for f\/ield theories based on the use of
Lie algebroids has been developed in recent works \cite{LMSV-2009,Ma-2004,Ma-2004b}.

In this work, we consider only the {\it multisymplectic} models
\cite{CM-2003,Go-91b,GIMMSY-mm,La-2000,MS-99},
f\/irst introduced by Tulczyjew and other authors
\cite{Gc-74,GS-73,Ki-73,KT-79}.
They arise from the study of multisymplectic manifolds
and their properties (see \cite{CIL-96a,CIL-96b} for general references,
and Appendix \ref{mf} for a~brief review);
in particular, those concerning the behavior of
multisymplectic Lagrangian and Hamiltonian systems.

The usual way of working with f\/ield theories
consists in stating their Lagrangian formalism
\cite{AA-80,BSF-88,CGR-2001,EMR-96,EMR-98,Gc-74,GMS-97,GS-73,Sa-89},
and jet bundles are the appropriate domain for doing so.
The construction of this formalism for regular and
singular theories is reviewed in Section~\ref{lf}.

The Hamiltonian description
presents dif\/ferent kinds of problems.
For instance, the choice of the multimomentum bundle for developing
the theory is not unique \cite{EMR-00a,EMR-00}, and dif\/ferent kinds
of Hamiltonian systems can be def\/ined, depending on this choice
and on the way of introducing the physical content (the ``Hamiltonian'')
\cite{ELMR-2007,EM-92,HK-01,HK-04,MPSW-2004,PR-2002}.
Here we present one of the most standard ways of def\/ining
Hamiltonian systems, which is based on using
{\it Hamiltonian sections}~\cite{CCI-91};
although this construction can also be done
taking {\it Hamiltonian densities}
\cite{CCI-91,GMS-97,MS-99,Sd-95}.
In particular, the construction of Hamiltonian systems
which are the Hamiltonian counterpart of Lagrangian systems
is carried out by using the {\it Legendre map} associated with the
Lagrangian system, and this problem has been studied
by dif\/ferent authors in the {\it $($hyper$)$ regular} case
\cite{CCI-91,Sa-89}, and in the
{\it singular} ({\it almost-regular}) case
\cite{GMS-97,LMM-96,Sd-95}. In Section~\ref{hf} we review
some of these constructions.

 Another subject of interest in the geometrical description of
 classical f\/ield theories concerns the f\/ield equations.
 In the multisymplectic models, both in the Lagrangian and Hamiltonian
 formalisms, these equations can be derived from a suitable
 variational principle: the so-called
 {\it Hamilton principle} in the Lagrangian formalism and
 {\it Hamilton--Jacobi principle} in the Hamiltonian formulation
 \cite{AA-80,ELMR-2007,EMR-96,EMR-00,Gc-74,GS-73}, and the
 f\/ield equations are usually written by using the
 multisymplectic form in order to characterize the critical
 sections which are solutions of the problem.
 In addition, these critical sections can be thought of as being
the integral manifolds of certain kinds of integrable multivector f\/ields or
Ehresmann connections, def\/ined in the bundles where the formalism is developed,
and satisfying a suitable geometric equation
which is the intrinsic formulation of the
systems of partial dif\/ferential equations locally describing the f\/ield
\cite{EMR-96,EMR-98,EMR-99b,LMM-96,Sa-89}.
All these aspects are discussed in Sections \ref{lf} and \ref{hf}
(furthermore, a quick review on multivector f\/ields and connections
is given in Appendix \ref{mvf}).
Moreover, multivector f\/ields are also used in order to state
generalized Poisson brackets in the Hamiltonian formalism
of f\/ield theories \cite{FPR-2003,Ka-95,Ka-97b,Ka-98,PR-2002}.

In ordinary mechanics there is also a unif\/ied formulation of
Lagrangian and Hamiltonian formalisms \cite{SR-83}, which is
based on the use of the {\it Whitney sum} of the tangent and
cotangent bundles (the {\it velocity} and {\it momentum phase
spaces} of the system).
This formalism has been generalized for non-autonomous mechanics
\cite{BEMR-2008,CMC-2002,GM-2005} and recently for classical f\/ield theories~\mbox{\cite{ELMMR-2004,LMM-2002}}.
The main features of this formulation are explained in Section~\ref{ulhf}.

Finally, an example showing the application of these formalisms
is analyzed in Section \ref{examp}. A~last section is devoted to make a discussion
about the current status on the research on dif\/ferent topics
concerning the multisymplectic approach to classical f\/ield theories.

We ought to point out that
there are also geometric frameworks for describing
the non-covariant or space-time formalism of f\/ield theories, where
the use of {\it Cauchy surfaces} is the fundamental tool
\cite{Go-91c,GIMMSY-mm2,LMS-2004}.
Nevertheless we do not consider these topics in this survey.

As a review paper, this work recovers results and contributions from
several previous papers, such as
\cite{CCI-91,ELMMR-2004,EMR-96,EMR-98,EMR-99b,EMR-00,HK-01,
LMM-96,LMM-2002,PR-2002},
among others.

In this paper, manifolds are real, paracompact,
  connected and $C^\infty$, maps are $C^\infty$, and sum over crossed repeated
  indices is understood.

\section{Lagrangian formalism}
\protect\label{lf}

\subsection{Lagrangian systems}

A {\it classical field theory}
is described by the following elements:
First, we have
the {\it configuration fibre bundle} $\pi\colon E\to M$,
with $\dim\,M=m$ and $\dim\,E=n+m$,
where $M$ is an oriented manifold with volume form
$\omega\in\df^m(M)$.
$\pi^1\colon J^1\pi\to E$ is the f\/irst-order jet bundle
of local sections of $\pi$, which is also a bundle over $M$
with projection
$\bar\pi^1=\pi\circ\pi^1\colon J^1\pi\longrightarrow M$,
and $\dim\,J^1\pi=nm+n+m$.
We denote by $(x^\nu,y^A,v^A_\nu)$
($\nu = 1,\ldots,m$; $A= 1,\ldots,n$)
natural coordinates in $J^1\pi$ adapted to the bundle structure
and such that
$\omega=\d x^1\wedge\cdots\wedge\d x^m\equiv\d^mx$.
Second, we give the
{\it Lagrangian density}, which is a $\bar\pi^1$-semibasic $m$-form on
$J^1\pi$ and hence it can be expressed as $\Lag =\lag (\bar\pi^{1*}\omega)$,
where $\lag\in\Cinfty(J^1\pi)$
is the {\it Lagrangian function} associated with $\Lag$ and $\omega$.

The bundle $J^1\pi$ is endowed with a canonical structure,
${\cal V}\in\df^1(J^1\pi)\otimes\Gamma (J^1\pi,{\rm V}(\pi^1 ))
\otimes\Gamma (J^1\pi,\bar\pi^{1^*}\Tan M)$,
which is called the {\it vertical endomorphism}
\cite{EMR-96,Gc-74,GS-73,Sa-89}
(here ${\rm V}(\pi^1)$ denotes the vertical subbundle
with respect to the projection $\pi^1$,
and $\Gamma(J^1\pi,{\rm V}(\pi^1))$ the set of sections
in the corresponding bundle).
Then the {\it Poincar\'e--Cartan $m$ and $(m+1)$-forms}
associated with~$\Lag$ are def\/ined as
\[
\Theta_{\Lag}:=\inn({\cal V})\Lag+\Lag\in\df^{m}(J^1\pi)
,\qquad
\Omega_{\Lag}:= -\d\Theta_{\Lag}\in\df^{m+1}(J^1\pi)  .
\]
We have the following local expressions
(where $\d^{m-1}x_\alpha\equiv\inn\left(\derpar{}{x^\alpha}\right)\d^mx$):
\begin{gather}
\Theta_{\Lag} = \derpar{\lag}{v^A_\nu}\d y^A\wedge\d^{m-1}x_\nu -
\left(\derpar{\lag}{v^A_\nu}v^A_\nu -\lag\right)\d^mx,
\nonumber \\
\Omega_{\Lag} =
-\frac{\partial^2\lag}{\partial v^B_\nu\partial v^A_\alpha}
\d v^B_\nu\wedge\d y^A\wedge\d^{m-1}x_\alpha
-\frac{\partial^2\lag}{\partial y^B\partial v^A_\alpha}\d y^B\wedge
\d y^A\wedge\d^{m-1}x_\alpha
\nonumber  \\
\phantom{\Omega_{\Lag} =}{} +
\frac{\partial^2\lag}{\partial v^B_\nu\partial v^A_\alpha}v^A_\alpha
\d v^B_\nu\wedge\d^mx  +
\left(\frac{\partial^2\lag}{\partial y^B\partial v^A_\alpha}v^A_\alpha
 -\derpar{\lag}{y^B}+
\frac{\partial^2\lag}{\partial x^\alpha\partial v^B_\alpha}
\right)\d y^B\wedge\d^mx   .
\label{omegal}
\end{gather}

\begin{definition}
$\ls$ is said to be a {\it Lagrangian system}.
The Lagrangian system and the Lagrangian function are
said to be {\it regular} if $\Omega_{\Lag}$ is a multisymplectic
$(m+1)$-form (i.e., $1$-non\-degenerate) \cite{CCI-91,EMR-96}. Elsewhere they are
 {\it singular} (or {\it non-regular}).
\end{definition}

The regularity condition is locally equivalent to
$\det\big(\frac{\partial^2\lag}
{\partial v^A_\alpha\partial v^B_\nu}(\bar y)\big)\not= 0$,
$\forall\, \bar y\in J^1\pi$.
We must point out that, in f\/ield theories, the notion of regularity
is not uniquely def\/ined (for other approaches see, for instance,
\cite{Be-84,De-77,De-78,Kr-87,KS-01a,KS-01b}).

\subsection{Lagrangian f\/ield equations}

The Lagrangian f\/ield equations can be derived from a variational principle.
In fact:

\begin{definition} \label{hvp}
 Let $\ls$ be a Lagrangian system.
 Let $\Gamma(M,E)$ be the set of sections of $\pi$. Consider the map
 \begin{gather*}
 {\bf L} \colon \ \Gamma(M,E) \longrightarrow \Real,
 \\
 \phantom{{\bf L} \colon}{} \
 \phi \mapsto \int_M(j^1\phi)^*\Theta_\Lag   ,
  \end{gather*}
where the convergence of the integral is assumed.
 The {\rm variational problem} for this Lagrangian system
 is the search of the critical (or
 stationary) sections of the functional ${\bf L}$,
 with respect to the variations of $\phi$ given
 by $\phi_t =\sigma_t\circ\phi$, where $\{\sigma_t\}$ is a
 local one-parameter group of any compact-supported
 $Z\in\vf^{{\rm V}(\pi)}(E)$
 (the module of $\pi$-vertical vector f\/ields in $E$), that is:
 \[
  \frac{\d}{\d t}\Big\vert_{t=0}\int_M\big(j^1\phi_t\big)^*\Theta_\Lag = 0   .
  \]
   This is the {\it Hamilton principle} of the Lagrangian formalism.
\end{definition}

 The Hamilton principle is equivalent
 to f\/ind a distribution ${\cal D}$ in $J^1\pi$ such that:
\begin{enumerate}\itemsep=0pt
\item
 ${\cal D}$ is $m$-dimensional.
\item
 ${\cal D}$ is $\bar\pi^1$-transverse.
\item
 ${\cal D}$ is integrable (that is, {\it involutive\/}).
\item
 The integral manifolds of ${\cal D}$ are the canonical liftings
to $J^1\pi$ of the critical sections of the Hamilton principle.
\end{enumerate}

A distribution
${\cal D}$ satisfying $1$ and $2$ is associated with a connection
in the bundle $\bar\pi^1\colon J^1\pi\to M$
(integrable if $3$ holds), whose local expression is
 \begin{equation}
\nabla=
 \d x^\mu\otimes \left(\derpar{}{x^\nu}+F_\nu^A\derpar{}{y^A}+
 G_{\nu\rho}^A\derpar{}{v_\rho^A}\right)   .
 \label{nablal}
 \end{equation}
Furthermore, these kinds of integrable distributions
and the corresponding connections are associated with classes of integrable
(i.e., non-vanishing, locally decomposable and involutive)
$\bar\pi^1$-transverse $m$-multivector f\/ields in $J^1\pi$
(see Appendix \ref{mvf}).
If $2$ holds, the local expression in natural coordinates of an element of one
of these classes is
\begin{equation}
{\cal X}=\bigwedge_{\nu=1}^m
 f\left(\derpar{}{x^\nu}+F_\nu^A\derpar{}{y^A}+
 G_{\nu\rho}^A\derpar{}{v_\rho^A}\right)   , \qquad
 \mbox{\rm ($f\in\Cinfty(J^1\pi)$ non-vanishing)}   .
 \label{locmvf1}
 \end{equation}
If, in addition, the integral sections are holonomic
(that is, they are canonical liftings
of sections of $\pi\colon E\to M$), then the integrable connections
and their associated classes of multivector f\/ields are called
{\it holonomic}. To be holonomic is equivalent
to be integrable and {\it semi-holonomic}, that is,
$F_\nu^A=v_\nu^A$ in the above local expressions.
Then:

\begin{theorem}\label{equivcs}
 Let $\ls$ be a Lagrangian system.
 The following assertions on a
 section $\phi\in\Gamma(M,E)$ are equivalent:
 \begin{enumerate}\itemsep=0pt
 \item[$1.$]
 $\phi$ is a critical section for the variational problem posed by
the Hamilton principle.
 \item[$2.$]
 $(j^1\phi)^*\inn (X)\Omega_\Lag= 0$, $\forall\, X\in\vf (J^1\pi)$ $($see {\rm \cite{GS-73})}.
  \item[$3.$]
 If $(U;x^\nu,y^A,v_\nu^A)$ is a natural system of
 coordinates in $J^1\pi$, then
$j^1\phi {=}\big(x^\nu ,y^A(x^\eta),\derpar{y^A}{x^\nu}(x^\eta)\big)$
 in $U$ satisfies the Euler--Lagrange equations  $($see {\rm \cite{EMR-96,GS-73})}
  \begin{equation}
 \derpar{\lag}{y^A}\circ j^1\phi-
\derpar{}{x^\mu}\left(\derpar{\lag}{v_\mu^A}\circ j^1\phi\right)= 0   .
 \label{ELeqs}
  \end{equation}
\item[$4.$]
 $j^1\phi$ is an integral section of a class of holonomic multivector fields
 $\{ {\cal X}_\Lag\}\subset\vf^m(J^1\pi)$ satisfying $($see {\rm \cite{EMR-98})}:
  \begin{equation}
 \inn ({\cal X}_\Lag)\Omega_\Lag=0 , \qquad
 \forall\,{\cal X}_\Lag\in\{ {\cal X}_\Lag\}   .
  \label{hlageq1}
 \end{equation}
\item[$5.$]
 $j^1\phi$ is an integral section of a holonomic connection
$\nabla_\Lag$ in $J^1\pi$ satisfying $($see {\rm \cite{LMM-96})}:
\begin{equation}
\inn(\nabla_\Lag)\Omega_\Lag=(m-1)\Omega_\Lag  .
\label{nablaeq}
\end{equation}
 \end{enumerate}
 \end{theorem}

\begin{proof}
See \cite{EMR-96,EMR-98,Gc-74,GS-73,LMM-96,Sa-89}.

($1 \Longleftrightarrow 2$)\quad
Let $Z\in\vf^{{\rm V}(\pi)}(E)$ be a compact-supported vector f\/ield,
and $V\subset M$ an open set such that
 $\partial V$ is a $(m-1)$-dimensional manifold
and that $\bar\tau({\rm supp}\,(Z))\subset V$. We
denote by $j^1Z\in\vf(J^1\pi)$ the canonical lifting of $Z$ to $J^1\pi$;
and, if $Z\in\vf^{{\rm V}(\pi)}(E)$, then $j^1Z\in\vf^{{\rm V}(\bar\pi^1)}(J^1\pi)$
(see~\cite{EMR-96} for the details).
Therefore
\begin{gather*}
\frac{d}{d t}\Big\vert_{t=0}\int_M(j^1\phi_t)^*\Theta_\Lag   =
\frac{d}{d t}\Big\vert_{t=0}\int_V(j^1\phi_t)^*\Theta_\Lag =
\frac{d}{d t}\Big\vert_{t=0}\int_V[j^1(\sigma_t\circ\phi)]^*\Theta_\Lag
\\
\hphantom{\frac{d}{d t}\Big\vert_{t=0}\int_M(j^1\phi_t)^*\Theta_\Lag}{} =
\frac{d}{d t}\Big\vert_{t=0}\int_V(j^1\phi)^*[(j^1\sigma_t)^*\Theta_\Lag]=
\int_V(j^1\phi)^*\left(\lim_{t\to 0}
\frac{(j^1\sigma_t)^*\Theta_\Lag-\Theta_\Lag}{t}\right)
\\
\hphantom{\frac{d}{d t}\Big\vert_{t=0}\int_M(j^1\phi_t)^*\Theta_\Lag}{}=
\int_V(j^1\phi)^*\Lie(j^1Z)\Theta_\Lag=
\int_V(j^1\phi)^*[\inn(j^1Z)\d\Theta_\Lag+\d\inn(j^1Z)\Theta_\Lag]
\\ \hphantom{\frac{d}{d t}\Big\vert_{t=0}\int_M(j^1\phi_t)^*\Theta_\Lag}{} =
-\int_V(j^1\phi)^*[\inn(j^1Z)\Omega_\Lag-\d\inn(j^1Z)\Theta_\Lag]
\\  \hphantom{\frac{d}{d t}\Big\vert_{t=0}\int_M(j^1\phi_t)^*\Theta_\Lag}{}=
-\int_V(j^1\phi)^*\inn(j^1Z)\Omega_\Lag+
\int_V\d[(j^1\phi)^*\inn(Z)\Theta_\Lag]
\\ \hphantom{\frac{d}{d t}\Big\vert_{t=0}\int_M(j^1\phi_t)^*\Theta_\Lag}{} =
-\int_V(j^1\phi)^*\inn(j^1Z)\Omega_\Lag+
\int_{\partial V}(j^1\phi)^*\inn(j^1Z)\Theta_\Lag\\
\hphantom{\frac{d}{d t}\Big\vert_{t=0}\int_M(j^1\phi_t)^*\Theta_\Lag}{}=
-\int_V(j^1\phi)^*\inn(j^1Z)\Omega_\Lag   ,
\end{gather*}
 as a consequence of Stoke's theorem and the hypothesis made on the
supports of the vertical f\/ields.
Thus, by the fundamental theorem of the variational calculus
we conclude that
$\frac{d}{d t}\big\vert_{t=0}\int_V(j^1\phi_t)^*\Theta_\Lag=0$
if, and only if, $(j^1\phi)^*\inn(j^1Z)\Omega_\Lag=0$,
for every compact-supported $Z\in\vf^{{\rm V}(\pi)}(E)$.
However, as compact-supported vector f\/ields generate locally the
$\Cinfty(E)$-module of vector f\/ields in $E$,
it follows that the last equality holds for every
$Z\in\vf^{{\rm V}(\pi)}(E)$.

Now, suppose $\phi\in\Gamma(M,E)$ is a critical section; that is,
$(j^1\phi)^*\inn (j^1Z)\Omega_{\Lag} = 0$, for every $Z\in\vf^{{\rm V}(\pi)}(E)$,
and consider $X\in\vf (J^1E)$, which can be written as
$X=X_\phi+X_v$ where $X_\phi$ is tangent to the image of $j^1\phi$ and
$X_v$ is $\bar\pi^1$-vertical,
both in the points of the image of $j^1\phi$.
However, $X_v=(X_v-j^1(\pi^1_*X_v))+j^1(\pi^1_*X_v)$,
where $j^1(\pi^1_*X_v)$ is understood as the prolongation of
a vector f\/ield which coincides with $\pi^1_*X_v$ on the image of $\phi$.
Observe that $\pi^1_*(X_v-j^1(\pi^1_*X_v))=0$ on the points
of the image of $j^1\phi$. Therefore
\[
(j^1\phi)^*\inn (X)\Omega_{\Lag} =
(j^1\phi)^*\inn (X_\phi)\Omega_{\Lag} +
(j^1\phi)^*\inn (X_v-j^1(\pi^1_*X_v))\Omega_{\Lag} +
(j^1\phi)^*\inn (j^1(\pi^1_*X_v))\Omega_{\Lag}   .
\]
However, $(j^1\phi)^*\inn (X_\phi)\Omega_{\Lag} =0$,
because $X_\phi$ is tangent to the image of $j^1\phi$, hence $\Omega_{\Lag}$
acts on linearly dependent vector f\/ields.
Nevertheless, $(j^1\phi)^*\inn (X_v-j^1(\pi^1_*X_v))\Omega_{\Lag} =0$,
because $X_v-j^1(\pi^1_*X_v)$ is $\pi^1$-vertical and $\Omega_{\Lag}$
vanishes on these vector f\/ields, when it is restricted to $j^1\phi$.
Therefore, as $\phi$ is stationary and $\pi^1_*X_v\in\vf^{{\rm V}(\pi)}(E)$, we have
\[
\int_M(j^1\phi)^*\inn (X)\Omega_{\Lag} =
\int_M(j^1\phi)^*\inn (j^1(\pi^1_*X_v))\Omega_{\Lag} =0  .
\]

The converse is a consequence of the f\/irst paragraph, since the condition
$(j^1\phi)^*\inn (X)\Omega_\Lag= 0$, $\forall\, X\in\vf (J^1\pi)$, holds,
in particular, for $j^1Z$, for every $Z\in\vf^{{\rm V}(\pi)} (E)$.

 ($2 \Leftrightarrow 3$)\quad
If $X=\alpha^\nu\derpar{}{x^\nu}+
 \beta^A\derpar{}{y^A}+\gamma_\nu^A\derpar{}{v_\nu^A}\in\vf(J^1\pi)$,
 taking into account the local expression~\eqref{omegal} of $\Omega_\Lag$, we have
 \begin{gather*}
 \inn (X)\Omega_\Lag  =
 (-1)^\eta\alpha^\nu
 \left[\frac{\partial^2\lag}{\partial v^B_\mu \partial v^A_\eta} \d v^B_\mu\wedge\d y^A\wedge\d^{m-2}x_{\eta\nu}+
 \frac{\partial^2\lag}{\partial y^B \partial v^A_\eta} \d y^B\wedge\d y^A\wedge\d^{m-2}x_{\eta\nu} \right.
 \\
 \left.
 \phantom{\inn (X)\Omega_\Lag  =}{}
 -\frac{\partial^2\lag}{\partial v^B_\mu \partial v^A_\eta}v^A_\eta \d v^B_\mu\wedge\d^{m-1}x_\nu
-\left(\frac{\partial^2\lag}{\partial y^B\partial v^A_\eta}v^A_\eta
 -\derpar{\lag}{y^B}+\frac{\partial^2\lag}{\partial x^\eta\partial v^B_\eta}\right)\d y^B \!\wedge\d^{m-1}x_\nu\right]
 \!
 \\
  \phantom{\inn (X)\Omega_\Lag  =}{}
 +\beta^A\left[\frac{\partial^2\lag}{\partial v^B_\mu\partial v^A_\eta}\d v^B_\mu\wedge\d^{m-1}x_\eta+
 \left( \frac{\partial^2\lag}{\partial y^A \partial v^B_\eta} - \frac{\partial^2\lag}{\partial y^B \partial v^A_\eta}\right)
 \d y^B\wedge\d^{m-1}x_\eta \right.
 \\
 \left.
  \phantom{\inn (X)\Omega_\Lag  =}{}
 +\left(\frac{\partial^2\lag}{\partial y^A\partial v^B_\eta}v^B_\eta
 -\derpar{\lag}{y^A}+\frac{\partial^2\lag}{\partial x^\eta\partial v^A_\eta}\right)\d^mx  \right]
 \\
  \phantom{\inn (X)\Omega_\Lag  =}{}
 + \gamma_\nu^A \left[
 -\frac{\partial^2\lag}{\partial v^A_\nu \partial v^B_\eta}\d y^B\wedge\d^{m-1}x_\eta
 +\frac{\partial^2\lag}{\partial v^A_\nu \partial v^B_\eta}v^B_\eta\d^mx \right]
 \end{gather*}
 but if $\phi=(x^\mu ,y^A(x^\eta))$, then
$j^1\phi =(x^\mu ,y^A(x^\eta),v^A(x^\eta))=
 \big( x^\mu ,y^A(x^\eta),\derpar{y^A}{x^\mu}(x^\eta)\big)$,
  and hence
  \begin{gather*}
 (j^1\phi)^*\inn (X)\Omega_\Lag   =
 (-1)^{\eta+\nu}\alpha^\eta\left[
 \derpar{}{x^\mu}\left(\derpar{\lag}{v_\mu^A}\circ j^1\phi\right)- \derpar{\lag}{y^A}\circ j^1\phi \right]
\derpar{(y^A \circ\phi)}{x^\eta} \d^mx
 \\
 \phantom{(j^1\phi)^*\inn (X)\Omega_\Lag   =}{}
+ \beta^A\left[\derpar{}{x^\mu}\left(\derpar{\lag}{v_\mu^A}\circ j^1\phi\right)- \derpar{\lag}{y^A}\circ j^1\phi\right]\d^mx    ,
 \end{gather*}
 and, as this holds for every $X\in\vf (J^1\pi)$,
 we conclude that $(j^1\phi)^*\inn (X)\Omega_{\Lag}=0$ if, and only if,
 the Euler--Lagrange equations (\ref{ELeqs}) hold for $\phi$.

 ($3 \Leftrightarrow 4$)\quad
  Using the local expressions (\ref{omegal}) of
 $\Omega_h$ and (\ref{locmvf1}) for ${\cal X}_\Lag$, and taking $f=1$
 as a~representative of the class $\{ {\cal X}_\Lag\}$, from the equation (\ref{hlageq1}),
 we obtain that
\begin{gather}
0  =  \big(F^B_\mu-v^B_\mu\big)\frac{\partial^2\lag}{\partial v^A_\nu\partial v^B_\mu}   ,
\label{eqs1}
\\
0  =  \derpar{\lag}{y^A}-\frac{\partial^2\lag}{\partial x^\mu\partial  v^A_\mu}-
\frac{\partial^2\lag}{\partial y^B\partial v^A_\mu}F^B_\mu-
\frac{\partial^2\lag}{\partial v^B_\nu\partial  v^A_\mu}G^B_{\nu\mu}+
\frac{\partial^2\lag}{\partial y^A\partial  v^B_\mu}\big(F^B_\mu-v^B_\mu\big)   ,
\label{eqs2}
\end{gather}
but, if $X_{\Lag}$ is holonomic, it is semiholonomic and then
$F^B_\mu=v^B_\mu$.
Therefore the equations (\ref{eqs1}) are identities,
and the equations (\ref{eqs2}) are
\begin{equation}
0=\derpar{\lag}{y^A}-\frac{\partial^2\lag}{\partial x^\mu\partial  v^A_\mu}-
\frac{\partial^2\lag}{\partial y^B\partial v^A_\mu}v^B_\mu-
\frac{\partial^2\lag}{\partial v^B_\nu\partial  v^A_\mu}G^B_{\nu\mu} \ .
\label{eqsG}
\end{equation}
Now, for a section $\phi=(x^\mu ,y^A(x^\eta))$, if
$  j^1\phi =\big( x^\mu ,y^A(x^\eta),\derpar{y^A}{x^\mu}(x^\eta)\big)$
is an integral section of $X_{\Lag}$, then
$G_{\nu\mu}^A=\frac{\partial^2y^A}{\partial x^\nu\partial x^\mu}$,
and the equations~(\ref{eqsG}) are equivalent to the Euler--Lagrange equations for $\phi$.

($3 \Leftrightarrow 5$)\quad
  The proof is like in the above item:
   using the local expressions (\ref{omegal}) of
 $\Omega_\Lag$ and~(\ref{nablal}) for $\nabla_\Lag$, we prove that the equation
 (\ref{nablaeq}) holds  for an integrable connection if, and only if, the
Euler--Lagrange equations (\ref{ELeqs}) hold for its integral sections.
\end{proof}

 {\it Semi-holonomic} (but not necessarily integrable)
 locally decomposable multivector f\/ields and connections
 which are solution to the Lagrangian equations (\ref{hlageq1}) and
(\ref{nablaeq}) respectively
 are called {\it Euler--Lagrange multivector fields} and {\it connections}
 for $\ls$.

If $\ls$ is regular, Euler--Lagrange $m$-multivector f\/ields and connections
exist in~$J^1\pi$, although they are not necessarily integrable.
If $\ls$ is singular, in the most favourable cases,
Euler--Lagrange multivector f\/ields and connections
only exist in some submanifold \mbox{$S\hookrightarrow J^1\pi$},
which can be obtained after applying a suitable constraint algorithm
(see~\cite{LMMMR-2005}).

\section{Hamiltonian formalism}
\protect\label{hf}

\subsection{Multimomentum bundles. Legendre maps}

As we have pointed out in the introduction,
the construction of the Hamiltonian formalism of f\/ield theories is more involved than
the Lagrangian formulation.
In fact, there are dif\/ferent bundles where the Hamiltonian formalism
can be developed (see, for instance, \cite{EMR-00a}, and references therein).
 Here we take one of the most standard choices.

First, ${\cal M}\pi\equiv\Lambda_2^m\Tan^*E$,
is the bundle of $m$-forms on
$E$ vanishing by the action of two $\pi$-vertical vector f\/ields
(so $\dim\, {\cal M}\pi=nm+n+m+1$),
and is dif\/feomorphic to the set ${\rm Af\/f}(J^1\pi,\Lambda^m\Tan^*M)$,
made of the af\/f\/ine maps from $J^1\pi$ to $\Lambda^m\Tan^*M$
(the multicotangent bundle of $M$ of order $m$~\cite{CIL-96b})
\mbox{\cite{CCI-91,EMR-00}}.
It is called the {\it extended multimomentum bundle},
and its canonical submersions are denoted
\[
\kappa\colon \ {\cal M}\pi\to E   ; \qquad
\bar\kappa=\pi\circ\kappa\colon \ {\cal M}\pi\to M  .
\]
As ${\cal M}\pi$ is a subbundle of $\Lambda^m\Tan^*E$,
then ${\cal M}\pi$ is endowed with a canonical form
$\Theta\in\df^m({\cal M}\pi)$ (the ``tautological form''),
which is def\/ined as follows:
let $(x,\alpha )\in\Lambda_2^m\Tan^*E $, with $x\in E$ and
$\alpha\in\Lambda_2^m\Tan_x^*E$; then,
for every $X_1,\ldots,X_m\in\Tan_{(x,\alpha)}({\cal M}\pi)$,
\[
\Theta ((x,\alpha );\moment{X}{1}{m}):=
\alpha (x;\Tan_{(x,\alpha)}\kappa(X_1),\ldots ,\Tan_{(q,\alpha)}\kappa(X_m))  .
\]
Then we def\/ine the multisymplectic form
 $\Omega:=-\d\Theta\in\df^{m+1}({\cal M}\pi)$.
They are known as the {\it multimomentum Liouville $m$ and $(m+1)$-forms}

If we introduce natural coordinates $(x^\nu,y^A,p^\nu_A,p)$ in ${\cal M}\pi$
adapted to the bundle $\pi\colon E\to M$,   and such that
$\omega=\d^mx$, the local expressions of these forms are
\[
 \Theta=p^\nu_A\d y^A\wedge\d^{m-1}x_\nu+p\d^mx
 , \qquad
 \Omega=-\d p^\nu_A\wedge\d y^A\wedge\d^{m-1}x_\nu-\d p\wedge\d^mx  .
 \]

Now we denote by $J^1\pi^*$ the quotient
${\cal M}\pi/\pi^*\Lambda^m\Tan^*M$, with $\dim\, J^1\pi^*=nm+n+m$.
We have the natural submersions
 \[
  \tau\colon \ J^1\pi^*\to E;\qquad
 \bar\tau=\pi\circ\tau\colon \ J^1\pi^*\to M   .
  \]
  Furthermore, the natural submersion $\mu\colon{\cal M}\pi\to J^1\pi^*$
endows ${\cal M}\pi$ with the structure of an af\/f\/ine bundle over $J^1\pi^*$,
with $(\pi\circ\tau)^*\Lambda^m\Tan^*M$ as the associated vector bundle.
$J^1\pi^*$ is usually called the {\it restricted multimomentum bundle}
associated with the bundle $\pi\colon E\to M$.

Natural coordinates in $J^1\pi^*$
(adapted to the bundle $\pi\colon E\to M$) are denoted by $(x^\nu,y^A,p^\nu_A)$.

\begin{definition}
Let $\ls$ be a Lagrangian system.
The {\it extended Legendre map} associated with $\Lag$,
 $\widetilde{{\cal F}\Lag}\colon J^1\pi\to {\cal M}\pi$, is
def\/ined by
 \[
  (\widetilde{{\cal F}\Lag}(\bar y))(\moment{Z}{1}{m}):=
 (\Theta_{\Lag})_{\bar y}(\moment{\bar Z}{1}{m})   ,
  \]
   where $\moment{Z}{1}{m}\in\Tan_{\pi^1(\bar y)}E$, and
 $\moment{\bar Z}{1}{m}\in\Tan_{\bar y}J^1\pi$ are such that
 $\Tan_{\bar y}\pi^1\bar Z_\alpha=Z_\alpha$.

The {\it restricted Legendre map} associated with $\Lag$ is
 ${\cal F}\Lag :=\mu\circ\widetilde{{\cal F}\Lag}\colon J^1\pi\to J^1\pi^*$.
\end{definition}

In natural coordinates we have:
  \begin{alignat*}{5}
& \widetilde{{\cal F}\Lag}^*x^\alpha = x^\alpha   , \qquad & &
 \widetilde{{\cal F}\Lag}^*y^A = y^A   , \qquad &&
 \widetilde{{\cal F}\Lag}^*p_A^\alpha =\derpar{\lag}{v^A_\alpha}
   , \qquad & &
 \widetilde{{\cal F}\Lag}^*p =\lag-v^A_\alpha\derpar{\lag}{v^A_\alpha}, &
 \\
& {\cal F}\Lag^*x^\alpha = x^\alpha   , \qquad & &
 {\cal F}\Lag^*y^A = y^A   , \qquad &&
 {\cal F}\Lag^*p_A^\alpha =\derpar{\lag}{v^A_\alpha} . &&&
  \end{alignat*}

Then, observe that
 $\widetilde{{\cal F}\Lag}^*\Theta=\Theta_{\Lag}$,
 and $\widetilde{{\cal F}\Lag}^*\Omega=\Omega_{\Lag}$.

\begin{definition}
$\ls$ is {\it regular} ({\it hyper-regular})
 if ${\cal F}\Lag$ is a local (global) dif\/feomorphism.
 Elsewhere it is {\it singular}.
(This def\/inition is equivalent to that given above.)

$\ls$ is {\it almost-regular} if
\begin{enumerate}\itemsep=0pt
\item
 ${\cal P}:={\cal F}\Lag (J^1\pi)$ is a closed submanifold of $J^1\pi^*$
 (natural embedding $\jmath_0\colon {\cal P}\hookrightarrow J^1\pi^*$).
\item
 ${\cal F}\Lag$ is a submersion onto its image.
\item
The f\/ibres ${\cal F}\Lag^{-1}({\cal F}\Lag (\bar y))$,
$\forall\, \bar y\in J^1\pi$,
 are connected submanifolds of $J^1\pi$.
\end{enumerate}
\end{definition}

\subsection{The (hyper)regular case}

In the Hamiltonian formalism of f\/ield theories,
there are dif\/ferent ways of introducing the physi\-cal information
(the ``Hamiltonian''). For instance, we can use connections in the
multimomentum bundles in order to obtain a covariant def\/inition of the
so-called {\it Hamiltonian densities} (see, for instance, \cite{CCI-91,GMS-97,MS-99,Sd-95}).

Nevertheless, the simplest way of def\/ining (regular) Hamiltonian systems in f\/ield theory
consists in considering the bundle $\bar\tau\colon J^1\pi^*\to M$
and then giving sections $h\colon J^1\pi^*\to{\cal M}\pi$ of the projection
 $\mu$, which are called {\it Hamiltonian sections}
and carry the physical information of the system.
Then we can def\/ine the dif\/ferentiable forms
\[
 \Theta_{h}:=h^*\Theta\in\df^m(J^1\pi^*)   ,\qquad
 \Omega_{h}:=-\d\Theta_{h}=h^*\Omega\in\df^{m+1}(J^1\pi^*)
\]
 which are the {\it Hamilton--Cartan $m$ and $(m+1)$ forms} of $J^1\pi^*$
 associated with the Hamiltonian section $h$.
 The couple $\hsjpi$ is said to be a {\it Hamiltonian system}.

 In a local chart of natural coordinates,
 a Hamiltonian section is specif\/ied by a {\it local Hamiltonian function}
 ${\rm h}\in\Cinfty (U)$, $U\subset J^1\pi^*$, such that
 $h(x^\nu,y^A,p^\nu_A)\equiv
 (x^\nu,y^A,p^\nu_A,p=-{\rm h}(x^\gamma,y^B,p_B^\eta))$. Then,
 the local expressions of the Hamilton--Cartan forms
 associated with $h$ are
\begin{gather}\label{omegah}
 \Theta_h = p_A^\nu\d y^A\wedge\d^{m-1}x_\nu -{\rm h}\d^mx
  , \qquad
 \Omega_h = -\d p_A^\nu\wedge\d y^A\wedge\d^{m-1}x_\nu +
 \d {\rm h}\wedge\d^mx.
 \end{gather}
 Notice that
$\Omega_h$ is $1$-nondegenerate; that is, a multisymplectic
form (as a simple calculation in coordinates shows).

Now we want to associate Hamiltonian systems to the Lagrangian ones.
First we consider the hyper-regular case
(the regular case is analogous, but working locally).

If $\ls$ is a hyper-regular Lagrangian system, then we have the diagram
 \[
 \begin{array}{ccccc}
\begin{picture}(15,52)(0,0)
\put(0,0){\mbox{$J^1\pi$}}
\end{picture}
&
\begin{picture}(65,52)(0,0)
 \put(20,27){\mbox{$\widetilde{{\cal F}\Lag}$}}
 \put(25,6){\mbox{${\cal F}\Lag$}}
 \put(0,7){\vector(2,1){65}}
 \put(0,4){\vector(1,0){65}}
\end{picture}
&
\begin{picture}(15,52)(0,0)
 \put(0,0){\mbox{$J^1\pi^*$}}
 \put(0,41){\mbox{${\cal M}\pi$}}
 \put(5,38){\vector(0,-1){25}}
 \put(-5,22){\mbox{$\mu$}}
 \put(10,13){\vector(0,1){25}}
 \put(15,22){\mbox{$h$}}
\end{picture}
\end{array}
 \]
 It is proved \cite{CCI-91} that
 $\tilde{\cal P}:=\widetilde{{\cal F}\Lag}(J^1\pi)$ is a
 1-codimensional imbedded submanifold of ${\cal M}\pi$
 ($\tilde\jmath_0\colon\tilde{\cal P}\hookrightarrow{\cal M}\pi$
denotes is the natural embedding),
which is transverse to $\mu$, and is dif\/feomorphic to  $J^1\pi^*$.
This dif\/feomorphism is $\mu^{-1}$, when $\mu$ is
 restricted to $\tilde{\cal P}$, and also coincides with the map
 $h:=\widetilde{{\cal F}\Lag}\circ{\cal F}\Lag^{-1}$,
 when it is restricted onto its image (which is just $\tilde{\cal P}$).
Thus $h$ and $\hsjpi$ are the {\it Hamiltonian section}
and the {\it Hamiltonian system associated with the
hyper-regular Lagrangian system $\ls$}, respectively.

 Locally, the Hamiltonian section
$h(x^\nu,y^A,p^\nu_A)=(x^\nu,y^A,p^\nu_A,p=-{\rm h}(x^\gamma,y^B,p_B^\gamma))$
 is specif\/ied by the {\it local Hamiltonian function}
 \[
 {\rm h}=p^\nu_A ({\cal F}\Lag^{-1})^*v_\nu^A-({\cal F}\Lag^{-1})^*\lag   .
\]
Then we have the local expressions (\ref{omegah}) for the
corresponding Hamilton--Cartan forms and, of course, ${\cal
F}\Lag^*\Theta_h=\Theta_{\Lag}$,
               and ${\cal F}\Lag^*\Omega_h=\Omega_{\Lag}$.

The Hamiltonian f\/ield equations can also be derived from a variational principle.
In fact:

 \begin{definition}
 Let $\hsjpi$ be a Hamiltonian system.
 Let $\Gamma(M,J^1\pi^*)$ be
 the set of sections of $\bar\tau$. Consider the map
  \begin{gather*}
 {\bf H} \colon \ \Gamma(M,J^1\pi^*) \longrightarrow \Real,\\
 \phantom{{\bf H} \colon}{} \ \psi \mapsto \int_M\psi^*\Theta_{h},
  \end{gather*}
 where the convergence of the integral is assumed.
 The {\rm variational problem} for this Hamiltonian system
 is the search for the critical (or
 stationary) sections of the functional ${\bf H}$,
 with respect to the variations of $\psi$ given
 by $\psi_t =\sigma_t\circ\psi$, where $\{\sigma_t\}$ is the
 local one-parameter group of any compact-supported
 $Z\in\vf^{{\rm V}(\bar\tau)}(J^1\pi^*)$
 ( the module of $\bar\tau$-vertical vector f\/ields in $J^1\pi^*$), that is:
 \[
  \frac{\d}{\d t}\Big\vert_{t=0}\int_M\psi_t^*\Theta_{h} = 0   .
  \]
   This is the so-called {\it Hamilton--Jacobi principle}
 of the Hamiltonian formalism.
 \label{hjvp}
 \end{definition}

 The Hamilton--Jacobi principle is equivalent
 to f\/ind distributions ${\cal D}$ of $J^1\pi^*$ such that:
\begin{enumerate}\itemsep=0pt
\item
 ${\cal D}$ is $m$-dimensional.
\item
 ${\cal D}$ is $\bar\tau$-transverse.
\item
 ${\cal D}$ is integrable (that is, {\it involutive\/}).
\item
 The integral manifolds of ${\cal D}$ are the critical sections of
 the Hamilton--Jacobi principle.
\end{enumerate}
As in the Lagrangian formalism,
${\cal D}$ are associated with classes of integrable and
 $\bar\tau$-transverse $m$-multivector f\/ields
 $\{ {\cal X}\}\subset\vf^m(J^1\pi^*)$ or,
 what is equivalent, with
 connections in the bundle  $\bar\pi\colon J^1\pi\to M$,
 whose expressions are
 \begin{gather}
{\cal X}  =  \bigwedge_{\nu=1}^m
 f\left(\derpar{}{x^\nu}+F_\nu^A\derpar{}{y^A}+
 G^\rho_{A\nu}\derpar{}{p^\rho_A}\right)  , \qquad
 \mbox{\rm ($f\in\Cinfty(J^1\pi^*)$ non-vanishing)},
 \label{locmvf2}  \\
\nabla  =
 \d x^\mu\otimes \left(\derpar{}{x^\mu}+F_\mu^A\derpar{}{y^A}+
 G^\rho_{A\mu}\derpar{}{p^\rho_A}\right)   .
 \label{locnablah}
  \end{gather}

Then we have:

\begin{theorem}\label{equics}
 The following assertions on a
 section $\psi\in\Gamma(M,J^1\pi^*)$ are equivalent:
 \begin{enumerate}\itemsep=0pt
 \item[$1.$]
 $\psi$ is a critical section for the variational problem posed by
the Hamilton--Jacobi principle.
 \item[$2.$]
 $\psi^*\inn (X)\Omega_{h}= 0$, $\forall\, X\in\vf (J^1\pi^*)$.
  \item[$3.$]
 If $(U;x^\nu,y^A,p_A^\nu)$ is a natural system of
 coordinates in $J^1\pi^*$, then $\psi$
 satisfies the  Hamilton--De Donder--Weyl equations in $U$
 \begin{equation}
 \derpar{(y^A\circ\psi)}{x^\nu}=
 \derpar{{\rm h}}{p^\nu_A}\circ\psi
,\qquad
 \derpar{(p_A^\nu\circ\psi)}{x^\nu}=
 - \derpar{{\rm h}}{y^A}\circ\psi .
 \label{HDWeqs}
 \end{equation}
  \item[$4.$]
 $\psi$ is an integral section of a class of integrable and
 $\bar\tau$-transverse multivector fields
 $\{ {\cal X}_h\}\subset\vf^m(J^1\pi^*)$ satisfying that
 \begin{equation}
 \inn ({\cal X}_h)\Omega_h=0 , \qquad
 \forall\,{\cal X}_h\in\{ {\cal X}_h\}  .
  \label{hameq1}
\end{equation}
\item[$5.$]
$\psi$ is an integral section of an integrable connection
$\nabla_h$ in $J^1\pi^*$ satisfying the equation
\begin{equation}
\inn(\nabla_h)\Omega_h=(m-1)\Omega_h  .
\label{nablaheq}
\end{equation}
 \end{enumerate}
 \end{theorem}

\begin{proof}
This proof is taken from \cite{ELMR-2007,EMR-99b}, and \cite{EMR-00}.

($1 \Leftrightarrow 2$)\quad
Let $Z\in\vf^{{\rm V}(\bar\tau)}(J^1\pi^*)$ be a compact-supported vector f\/ield,
and $V\subset M$ an open set such that
 $\partial V$ is a $(m-1)$-dimensional manifold
and that $\bar\tau({\rm supp}\,(Z))\subset V$. Then
\begin{gather*}
\frac{d}{d t}\Big\vert_{t=0}\int_M\psi_t^*\Theta_h   =
\frac{d}{d t}\Big\vert_{t=0}\int_V\psi_t^*\Theta_h =
\frac{d}{d t}\Big\vert_{t=0}\int_V\psi^*(\sigma_t^*\Theta_h) =
\int_V\psi^*\left(\lim_{t\to 0}
\frac{\sigma_t^*\Theta_h-\Theta_h}{t}\right)
\\
\hphantom{\frac{d}{d t}\Big\vert_{t=0}\int_M\psi_t^*\Theta_h}{} =
\int_V\psi^*\Lie(Z)\Theta_h =
\int_V\psi^*(\inn(Z)\d\Theta_h+\d\inn(Z)\Theta_h)
\\  \hphantom{\frac{d}{d t}\Big\vert_{t=0}\int_M\psi_t^*\Theta_h}{}=
-\int_V\psi^*(\inn(Z)\Omega_h-\d\inn(Z)\Theta_h)=
-\int_V\psi^*\inn(Z)\Omega_h+\int_V\d[\psi^*\inn(Z)\Theta_h]
\\ \hphantom{\frac{d}{d t}\Big\vert_{t=0}\int_M\psi_t^*\Theta_h}{} =
-\int_V\psi^*\inn(Z)\Omega_h+
\int_{\partial V}\psi^*\inn(Z)\Theta_h=
-\int_V\psi^*\inn(Z)\Omega_h   ,
\end{gather*}
as a consequence of Stoke's theorem and the hypothesis made on the
supports of the vertical f\/ields.
Thus, by the fundamental theorem of the variational calculus
we conclude that
$\frac{d}{d t}\big\vert_{t=0}\int_V\psi_t^*\Theta_h=0$
if, and only if, $\psi^*\inn(Z)\Omega_h=0$,
for every compact-supported $Z\in\vf^{{\rm V}(\bar\tau)}(J^1\pi^*)$.
However, as compact-supported vector f\/ields generate locally the
$\Cinfty(J^1\pi^*)$-module of vector f\/ields in $J^1\pi^*$,
it follows that the last equality holds for every
$Z\in\vf^{{\rm V}(\bar\tau)}(J^1\pi^*)$.

Now, if ${\rm p}\in{\rm Im}\,\psi$, then
$\Tan_{\rm p}J^1\pi^*=
{\rm V}_{\rm p}(\bar\tau)\oplus\Tan_{\rm p}({\rm Im}\,\psi)$.
So if $X\in\vf(J^1\pi^*)$, then
\[
X_{\rm p}=(X_{\rm p}-\Tan_{\rm p}(\psi\circ\bar\tau)(X_{\rm p}))+
\Tan_{\rm p}(\psi\circ\bar\tau)(X_{\rm p})\equiv
X^V_{\rm p}+X^{\psi}_{\rm p}   ,
\]
and therefore
\[
\psi^*\inn(X)\Omega_h=
\psi^*\inn(X^V)\Omega_h+
\psi^*\inn(X^{\psi})\Omega_h=
\psi^*\inn(X^{\psi})\Omega_h=0   ,
\]
since $\psi^*\inn(X^V)\Omega_h=0$, by the conclusion in the above paragraph.
Furthermore,
$X^{\psi}_{\rm p}\in\Tan_{\rm p}({\rm Im}\,\psi)$,
and $\dim\,({\rm Im}\,\psi)=m$, being
$\Omega_h\in\df^{m+1}(J^1\pi^*)$.
Hence we conclude that
 $\psi^*\inn (X)\Omega_h= 0$, for every
 $X\in\vf (J^1\pi^*)$.

The converse is obvious taking into account
the reasoning of the f\/irst paragraph, since the condition
$\psi^*\inn (X)\Omega_{h}= 0$, $\forall\, X\in\vf (J^1\pi^*)$, holds,
in particular, for every $Z\in\vf^{{\rm V}(\bar\tau)} (J^1\pi^*)$.

   ($2 \Leftrightarrow 3$)\quad
If $ X=\alpha^\nu\derpar{}{x^\nu}+
 \beta^A\derpar{}{y^A}+\gamma^\nu_A\derpar{}{p^\nu_A}\in\vf(J^1\pi^*)$,
 taking into account the local expression~(\ref{omegah}) of $\Omega_h$, we have
 \begin{gather*}
 \inn (X)\Omega_h  =
 (-1)^\eta\alpha^\eta\left(\d p^\nu_A\wedge\d y^A\wedge\d^{m-2}x_{\eta\nu}-
 \derpar{{\rm h}}{p^\nu_A} \d p^\nu_A \wedge\d^{m-1}x_\eta\right)
 \\  \phantom{\inn (X)\Omega_h  =}{}+
 \beta^A\left(\d p^\nu_A\wedge\d^{m-1}x_\nu +
 \derpar{{\rm h}}{y^A}\d^mx \right) +
 \gamma^\nu_A \left(-\d y^A\wedge\d^{m-1}x_\nu +
 \derpar{{\rm h}}{p^\nu_A}\wedge\d^mx\right)
 \end{gather*}
 but if $\psi =(x^\nu ,y^A(x^\eta ),p^\nu_A(x^\eta ))$, then
\begin{gather*}
 \psi^*\inn (X)\Omega_h   =
 (-1)^{\eta+\nu}\alpha^\eta\left(\derpar{(y^A\circ\psi)}{x^\nu}-
 \derpar{{\rm h}}{p^\nu_A}\Big\vert_\psi\right)
\derpar{(p^\nu_A\circ\psi)}{x^\eta} \d^mx
 \\ \phantom{\psi^*\inn (X)\Omega_h   =}{}
+ \beta^A\left(\derpar{(p^\nu_A\circ\psi)}{x^\nu}+
 \derpar{{\rm h}}{y^A}\Big\vert_\psi\right)\d^mx +
 \gamma^\nu_A\left(-\derpar{(y^A\circ\psi)}{x^\nu}+
 \derpar{{\rm h}}{p^\nu_A}\Big\vert_\psi \right)\d^mx   ,
 \end{gather*}
 and, as this holds for every $X\in\vf (J^1\pi^*)$,
 we conclude that $\psi^*\inn (X)\Omega_h=0$ if, and only if,
 the Hamilton--De Donder--Weyl equations (\ref{HDWeqs}) hold for $\psi$.

($3 \Leftrightarrow 4$)\quad
  Using the local expressions (\ref{omegah}) of
 $\Omega_h$ and (\ref{locmvf2}) for ${\cal X}_h$, and taking $f=1$
 as a~representative of the class $\{ {\cal X}_h\}$, the equation (\ref{hameq1}),
 in coordinates, is
 \[
 F^A_\nu = \derpar{{\rm h}}{p_A^\nu},\qquad
G^\nu_{A\nu}= -\derpar{{\rm h}}{y^A} .
\]
This result allows us to assure the local existence of (classes of) multivector f\/ields
satisfying the desired conditions. The corresponding global solutions
are then obtained using a partition of unity subordinated
to a covering of $J^1\pi^*$ made of local natural charts.
Now, if $\psi(x) =(x^\nu ,y^A(x^\gamma),p^\nu_A(x^\gamma))$
is an integral section of ${\cal X}_h$, then
 \[
  \derpar{(y^A\circ\psi)}{x^\nu}=F^A_\nu\circ\psi , \qquad
 \derpar{(p^\rho_A\circ\psi)}{x^\nu}=G^\rho_{A\nu}\circ\psi   .
 \]
 Thus, combining both expressions we obtain
the Hamilton--De Donder--Weyl equations (\ref{HDWeqs}) for~$\psi$.

($3 \Leftrightarrow 5$)\quad
  The proof is like in the above item:
   using the local expressions (\ref{omegah}) of
 $\Omega_h$ and (\ref{locnablah}) for $\nabla_h$, we prove that the equation
 (\ref{nablaheq}) holds  for an integrable connection if, and only if, the
Hamilton--De Donder--Weyl equations (\ref{HDWeqs}) hold for its integral sections.
\end{proof}

The $\bar\tau$-transverse locally decomposable multivector f\/ields
 and connections which are solution to the Hamiltonian equations (\ref{hameq1}) and
(\ref{nablaheq}) respectively (but not necessarily integrable)
 are called {\it Hamilton--De Donder--Weyl multivector fields} and
{\it connections} for $\hsjpi$.

Hence, the existence of Hamilton--De Donder--Weyl multivector f\/ields and connections
for $\hsjpi$ is assured, although they are not necessarily integrable.

Finally, we can establish the equivalence between the
Lagrangian and Hamiltonian formalisms in the hyper-regular case:

\begin{theorem} \emph{(equivalence theorem for sections)} Let $\ls$ be a
hyper-regular Lagrangian system,
and $ \hsjpi$ the associated Hamiltonian system.

If a section $\phi\in\Gamma(M,E)$ is a solution to the
Lagrangian variational problem $($Hamilton principle$)$, then the
section $\psi={\cal F}\Lag\circ j^1\phi\in\Gamma(M,J^1\pi^*)$
is a solution to the Hamiltonian
variational problem $($Hamilton--Jacobi principle$)$.

Conversely, if $\psi\in\Gamma(M,J^1\pi^*)$ is a solution to the
Hamiltonian variational problem, then the section
$\phi=\tau\circ\psi\in\Gamma(M,E)$ is a solution to the
Lagrangian variational problem.
\label{equiv1}
\end{theorem}

\begin{proof}
This proof is taken from \cite{EMR-99b} and \cite{EMR-00}.

Bearing in mind the diagram
\[
\begin{array}{ccc}
J^1\pi &
\begin{picture}(135,10)(0,0)
\put(63,6){\mbox{${\cal F}\Lag$}} \put(0,3){\vector(1,0){135}}
\end{picture}
& J^1\pi^*
\\ &
\begin{picture}(135,100)(0,0)
\put(34,82){\mbox{$\pi^1$}} \put(90,82){\mbox{$\tau$}}
\put(30,55){\mbox{$j^1\phi$}} \put(96,55){\mbox{$\psi$}}
\put(75,30){\mbox{$\pi$}} \put(55,30){\mbox{$\phi$}}
\put(63,55){\mbox{$E$}} \put(65,0){\mbox{$M$}}
\put(0,100){\vector(3,-2){55}} \put(135,100){\vector(-3,-2){55}}
\put(55,13){\vector(-2,3){55}} \put(80,13){\vector(2,3){55}}
\put(65,13){\vector(0,1){30}} \put(71,43){\vector(0,-1){30}}
\end{picture} &
\end{array}
\]
if $\phi$ is a solution to the Lagrangian
variational problem then $(j^1\phi)^*\inn (X)\Omega_{\Lag}=0$, for
every $X\in\vf (J^1\pi)$ (Theorem \ref{equivcs}, item 2); therefore, as
${\cal F}\Lag$ is a local dif\/feomorphism,
 \begin{gather*}
  0 = (j^1\phi)^*\inn
(X)\Omega_{\Lag}= (j^1\phi)^*\inn (X)({\cal F}\Lag^*\Omega_h)
\\ \phantom{0}{} =
(j^1\phi)^*{\cal F}\Lag^*(\inn ({\cal F}\Lag_*^{-1}X)\Omega_h)=
({\cal F}\Lag\circ j^1\phi)^*\inn (X')\Omega_h)   ,
 \end{gather*}
 which holds for every
$X'\in\vf (J^1\pi^*)$ and thus, by the item 2 of Theorem~\ref{equics},
 $\psi\equiv {\cal F}\Lag\circ j^1\phi$
 is a~solution to the Hamiltonian variational problem.

Conversely, let $\psi\in\Gamma(M,J^1\pi^*)$ be a solution to the
Hamiltonian variational problem. Reversing the above reasoning we
obtain that $({\cal F}\Lag^{-1}\circ\psi)^*\inn
(X)\Omega_{\Lag}=0$, for every $X\in\vf (J^1\pi)$, and hence
$\sigma\equiv {\cal F}\Lag^{-1}\circ\psi\in\Gamma(M,J^1E)$ is a
critical section for the Lagrangian variational problem. Then, as
we are in the hyper-regular case, $\sigma$ must be an holonomic
section, $\sigma =j^1\phi$ \cite{EMR-98,LMM-96,Sa-89},
 and since the above diagram is commutative,
$\phi =\tau^1\circ\psi\in\Gamma(M,E)$.
\end{proof}

The equivalence between the Lagrangian and the Hamiltonian formalisms can be
stated also in terms of multivector f\/ields and connections (see~\cite{EMR-99b}).

\subsection{The almost-regular case}

Now, consider the almost-regular case.
Let $\tilde{\cal P}:=\widetilde{{\cal F}\Lag}(J^1\pi)$,
${\cal P}:={\cal F}\Lag(J^1\pi)$
(the natural projections are denoted by $\tau_0^1\colon{\cal P}\to E$ and
$\bar\tau_0^1:=\pi\circ\tau_0^1\colon{\cal P}\to M$),
and  assume that ${\cal P}$ is a f\/ibre bundle over $E$ and $M$.
Denote by $\tilde\jmath_0\colon\tilde{\cal P}\hookrightarrow{\cal M}\pi$
the natural imbedding, and by
$\widetilde{{\cal F}\Lag}_0$ and ${\cal F}\Lag_0$
the restrictions of $\widetilde{{\cal F}\Lag}$ and ${\cal F}\Lag$
to their images, respectively. So, we have the diagram
 \[
 \begin{array}{cccc}
\begin{picture}(20,52)(0,0)
\put(0,0){\mbox{$J^1\pi$}}
\end{picture}
&
\begin{picture}(65,52)(0,0)
 \put(7,28){\mbox{$\widetilde{{\cal F}\Lag}_0$}}
 \put(24,7){\mbox{${\cal F}\Lag_0$}}
 \put(0,7){\vector(2,1){65}}
 \put(0,4){\vector(1,0){65}}
\end{picture}
&
\begin{picture}(90,52)(0,0)
 \put(5,0){\mbox{${\cal P}$}}
 \put(5,42){\mbox{$\tilde{\cal P}$}}
 \put(5,13){\vector(0,1){25}}
 \put(10,38){\vector(0,-1){25}}
 \put(-10,22){\mbox{$h_{\cal P}$}}
 \put(12,22){\mbox{$\tilde\mu$}}
 \put(30,45){\vector(1,0){55}}
 \put(30,4){\vector(1,0){55}}
 \put(48,12){\mbox{$\jmath_0$}}
 \put(48,33){\mbox{$\tilde\jmath_0$}}
 \end{picture}
&
\begin{picture}(15,52)(0,0)
 \put(0,0){\mbox{$J^1\pi^*$}}
 \put(0,41){\mbox{${\cal M}\pi$}}
 \put(10,38){\vector(0,-1){25}}
 \put(0,22){\mbox{$\mu$}}
\end{picture}
\\
& &
\begin{picture}(90,35)(0,0)
 \put(10,35){\vector(1,-1){35}}
 \put(5,11){\mbox{$\bar\tau_0$}}
 \put(90,11){\mbox{$\bar\tau$}}
 \put(100,35){\vector(-1,-1){35}}
\end{picture}
 &
\\
& & \qquad M &
 \end{array}
 \]
 Now, it can be proved that the $\mu$-transverse submanifold
$\tilde{\cal P}$ is dif\/feomorphic to ${\cal P}$ \cite{LMM-96}.
 This dif\/feomorphism is denoted
 $\tilde\mu\colon\tilde{\cal P}\to{\cal P}$,
 and it is just the restriction of the projection $\mu$ to $\tilde{\cal P}$.
 Then, taking $h_{\cal P}:=\tilde\mu^{-1}$,
 we def\/ine the Hamilton--Cartan forms
\[
 \Theta^0_h=(\tilde\jmath_0\circ h_{\cal P})^*\Theta\in\df^m({\cal P}) ,\qquad
 \Omega^0_h=
 -\d\Theta^0_h(\tilde\jmath_0\circ h_{\cal P})^*\Omega\in\df^{m+1}({\cal P})  ,
 \]
  which verify that
 ${\cal F}\Lag_0^*\Omega^0_h=\Omega_{\Lag}$.
 Then $h_{\cal P}$ is also called a {\it Hamiltonian section}, and
 $\hsjpio$ is the {\it Hamiltonian system}
 associated with the almost-regular Lagrangian system $\ls$.
In general, $\Omega_h^0$ is a pre-multisymplectic form
and $\hsjpio$ is the {\it Hamiltonian system associated with the
almost-regular Lagrangian system $\ls$}.

In this framework,
the {\it Hamilton--Jacobi principle} for $ \hsjpio$ is stated like above,
and the critical sections $\psi_0\in\Gamma(M,{\cal P})$
can be characterized in an analogous way than in Theorem
\ref{equics}.

If $\Omega_h^0$ is a pre-multisymplectic form,
Hamilton--De Donder--Weyl multivector vector f\/ields and connections
only exist, in the most favourable cases,
in some submanifold $S\hookrightarrow J^1\pi$,
and they are not necessarily integrable.
As in the Lagrangian case, $S$ can be obtained after applying the
suitable constraint algorithm \cite{LMMMR-2005}.
Then, the equivalence theorem follows in
an analogous way than above.

It is important to point out that the analysis of the Hamiltonian description of non-regular
f\/ield theories is far to be completed and, in fact, there is a lot of  topics
under discussion. For instance, there are some kinds of singular Lagrangian systems
for which the construction of the associated Hamiltonian formalism (following the
procedure that we have presented here) is ambiguous and,
in order to overcome this trouble, a dif\/ferent notion of regularity must be done,
which involve the use of {\it Lepagean forms} \cite{Kr-87, KS-01a, KS-01b}.
Neverthelees, the analysis of this and other problems exceeds the scope
of this work.

\section[Unified Lagrangian-Hamiltonian formalism]{Unif\/ied Lagrangian--Hamiltonian formalism}
\protect\label{ulhf}

\subsection{Geometric framework}

The {\it extended} and the
{\it restricted jet-multimomentum bundles} are
\[
{\cal W}:= J^1\pi \times_E {\cal M}\pi  , \qquad
{\cal W}_r := J^1\pi\times_E J^1\pi^*   ,
\]
with natural coordinates $(x^{\alpha},y^A,v^A_{\alpha},p_A^{\alpha},p)$
and $(x^{\alpha},y^A,v^A_{\alpha},p_A^{\alpha})$.
We have natural projections (submersions)
$\mu_{\cal W}\colon{\cal W}\to{\cal W}_r$, and
\begin{alignat}{5}
&\rho_1\colon{\cal W}\to J^1\pi ,\qquad &&
\rho_2\colon{\cal W}\to {\cal M}\pi ,\qquad &&
\rho_E\colon{\cal W}\to E ,\qquad &&
\rho_M\colon{\cal W}\to M, & \nonumber
\\
& \rho_1^r\colon{\cal W}_r\to J^1\pi ,\qquad &&
\rho_2^r\colon{\cal W}_r\to J^1\pi^* ,\qquad &&
\rho_E^r\colon{\cal W}_r\to E ,\qquad &&
\rho_M^r\colon{\cal W}_r\to M. &
\label{project}
\end{alignat}

\begin{definition}
The {\it coupling $m$-form} in ${\cal W}$, denoted by
${\cal C}$, is an $m$-form along $\rho_M$
which is def\/ined as follows:
for every $\bar y\in J_y^1E$, with $\bar\pi^1(\bar y)=\pi(y)=x\in E$,
and ${\bf p}\in{\cal M}_y\pi$, let $w\equiv (\bar y,{\bf p})\in{\cal W}_y$, then
\[
{\cal C}(w):=(\Tan_x\phi)^*{\bf p} ,
\]
where $\phi\colon M\to E$ satisf\/ies that $j^1\phi (x) = \bar y$.
Then, we denote by $\hat{\cal C}\in\df^m({\cal W})$ the
$\rho_M$-semibasic form associated with ${\cal C}$.

The {\it canonical $m$-form}
$\Theta_{\cal W}\in\df^m({\cal W})$ is def\/ined as
$\Theta_{\cal W}:=\rho_2^*\Theta$, and is $\rho _E$-semibasic.
The {\it canonical $(m+1)$-form} is the pre-multisymplectic form
$\Omega_{\cal W}:=-\d\Theta_{\cal W}=\rho_1^*\Omega\in\df^{m+1}({\cal W})$.
\label{coupling}
\end{definition}

There exists $\hat C\in\Cinfty ({\cal W})$ such that
$\hat{\cal C}=\hat C(\rho_M^*\omega)$, and
$\hat{\cal C}(w)=(p+p_A^\alpha v^A_\alpha)\d^mx$.

Local expressions of $\Theta_{\cal W}$ and $\Omega_{\cal W}$ are the same than for
$\Theta$ and $\Omega$.

Let $\hat\Lag:=\rho_1^*\Lag\in\df^m({\cal W})$,
and $\hat\Lag=\hat L(\rho_M^*\omega)$,
with $\hat L=\rho_1^*L\in\Cinfty ({\cal W})$.
We def\/ine the {\it Hamiltonian submanifold}
$\jmath_0\colon{\cal W}_0\hookrightarrow{\cal W}$ by
\[
{\cal W}_0:=\{ w\in{\cal W}\ | \ \hat\Lag(w)=\hat{\cal C}(w) \}  .
\]
The constraint function def\/ining ${\cal W}_0$ is
\[
\hat C-\hat L=p + p_A^\alpha v^A_\alpha-\hat L\big(x^\nu,y^B,v^B_\nu\big)=0  .
\]
There are projections which are the restrictions to ${\cal W}_0$
of the projections (\ref{project}), as it
is shown in the following diagram:
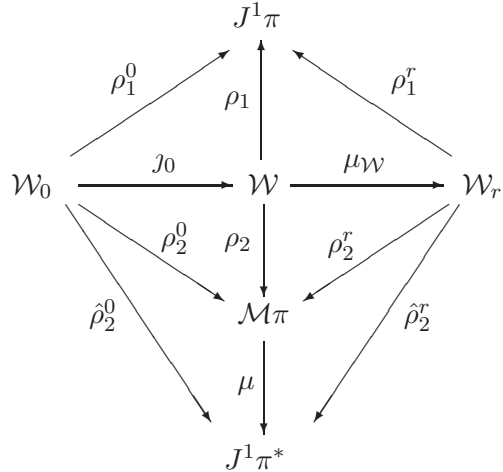
\[
\begin{array}{ccc}
& \begin{picture}(135,20)(0,0)
\put(58,5){\mbox{${J^1\pi}$}}
\end{picture} &
\\
& \begin{picture}(135,30)(0,0)
\put(12,16){\mbox{$\rho_1^0$}}
\put(-3,-10){\vector(3,2){60}}
\put(55,10){\mbox{$\rho_1$}}
\put(69,-11){\vector(0,1){45}}
\put(141,-10){\vector(-3,2){60}}
\put(118,16){\mbox{$\rho_1^r$}}
\end{picture}
\\
{\cal W}_0 &
\begin{picture}(135,20)(0,0)
\put(28,10){\mbox{$\jmath_0$}}
\put(0,3){\vector(1,0){58}}
\put(64,0){\mbox{${\cal W}$}}
\put(100,10){\mbox{$\mu_{\cal W}$}}
\put(80,3){\vector(1,0){58}}
\end{picture}
&  {\cal W}_r
\\ &
\begin{picture}(135,100)(0,0)
\put(31,84){\mbox{$\rho_2^0$}}
\put(55,84){\mbox{$\rho_2$}}
\put(94,82){\mbox{$\rho_2^r$}}
\put(4,55){\mbox{$\hat\rho_2^0$}}
 \put(123,55){\mbox{$\hat\rho_2^r$}}
 \put(60,30){\mbox{$\mu$}}
\put(60,55){\mbox{${\cal M}\pi$}}
 \put(55,0){\mbox{$J^1\pi^*$}}
\put(70,100){\vector(0,-1){35}}
\put(0,100){\vector(3,-2){55}}
 \put(140,100){\vector(-3,-2){55}}
\put(-5,100){\vector(2,-3){55}}
 \put(144,100){\vector(-2,-3){55}}
\put(70,48){\vector(0,-1){35}}
\end{picture} &
\end{array}
\]
$(x^{\alpha},y^A,v^A_\alpha,p_A^\alpha)$ are local coordinates in ${\cal W}_0$, and
\begin{alignat*}{3}
& \rho_1^0(x^{\alpha},y^A,v^A_\alpha,p_A^\alpha)=(x^{\alpha},y^A,v^A_\alpha),\qquad &&
\jmath_0(x^{\alpha},y^A,v^A_\alpha,p_A^\alpha)=
   (x^{\alpha},y^A,v^A_\alpha,p_A^\alpha,L-v^A_\alpha p_A^\alpha), & \\
& \hat\rho_2^0(x^{\alpha},y^A,v^A_\alpha,p_A^\alpha)=(x^{\alpha},y^A,p_A^\alpha)
,\qquad &&
\rho_2^0(x^{\alpha},y^A,v^A_\alpha,p_A^\alpha)=
(x^{\alpha},y^A,p_A^\alpha,L-v^A_\alpha p_A^\alpha) .&
\end{alignat*}

It is proved that
${\cal W}_0$ is a $1$-codimensional $\mu_{\cal W}$-transversal submanifold
of ${\cal W}$, dif\/feomorphic to~${\cal W}_r$.
As a consequence,
${\cal W}_0$ induces a {\it Hamiltonian section} of
$\mu_{\cal W}$,
$\hat h\colon{\cal W}_r\to{\cal W}$,
which is locally  specif\/ied by
giving the local {\it Hamiltonian function}
$\hat H= -\hat L+p_A^\alpha v^A_\alpha$; that is,
$\hat h(x^\alpha,y^A,v^A_\alpha,p^\alpha_A)=
(x^\alpha,y^A,v^A_\alpha,p^\alpha_A,-\hat H)$.
 From $\hat h$ we recover a Hamiltonian section
 $\tilde h\colon{\cal P}\to{\cal M}\pi$ def\/ined by
  $\tilde h([{\bf p}])=(\rho_2\circ\hat h)[(\rho_2^r)^{-1}(\jmath([{\bf p}]))]$,
$\forall\, [{\bf p}]\in{\cal P}$. (See the diagram.)
 
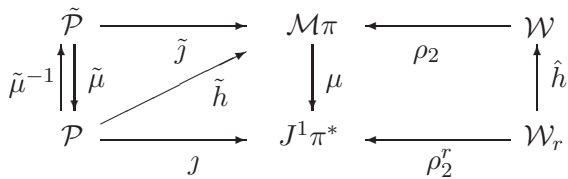
\[
 \begin{picture}(90,52)(0,0)
 \put(5,0){\mbox{${\cal P}$}}
 \put(5,42){\mbox{$\tilde{\cal P}$}}
 \put(5,13){\vector(0,1){25}}
 \put(10,38){\vector(0,-1){25}}
 \put(-15,20){\mbox{$\tilde\mu^{-1}$}}
 \put(15,22){\mbox{$\tilde\mu$}}
 \put(20,45){\vector(1,0){55}}
 \put(20,2){\vector(1,0){55}}
 \put(20,8){\vector(2,1){55}}
 \put(53,-9){\mbox{$\jmath$}}
 \put(48,33){\mbox{$\tilde\jmath$}}
 \put(62,15){\mbox{$\tilde h$}}
 \end{picture}
\begin{picture}(90,52)(0,0)
 \put(-3,0){\mbox{$J^1\pi^*$}}
 \put(0,42){\mbox{${\cal M}\pi$}}
 \put(10,38){\vector(0,-1){25}}
 \put(15,22){\mbox{$\mu$}}
 \put(85,45){\vector(-1,0){55}}
 \put(85,2){\vector(-1,0){55}}
 \put(53,-9){\mbox{$\rho_2^r$}}
 \put(48,33){\mbox{$\rho_2$}}
 \end{picture}
\begin{picture}(15,52)(0,0)
 \put(0,0){\mbox{${\cal W}_r$}}
 \put(0,41){\mbox{${\cal W}$}}
 \put(5,13){\vector(0,1){25}}
 \put(10,22){\mbox{$\hat h$}}
\end{picture}
 \]
 (For hyper-regular systems we have
 $\tilde{\cal P}={\cal M}\pi$ and ${\cal P}=J^1\pi^*$.)

We def\/ine the forms
$\Theta_0:=\jmath_0^*\Theta_{\cal W} = \rho _2^{0*} \Theta \in\df^m({\cal W}_0)$, and
$\Omega_0:=\jmath_0^*\Omega_{\cal W} =\rho _2^{0*} \Omega\in\df^{m+1}({\cal W}_0)$,
whose local expressions are
\begin{gather*}
\Theta_0 = (L-p_A^\alpha v^A_\alpha)\d^mx+
p_A^\alpha\d y^A\wedge\d^{m-1}x_\alpha,
\nonumber \\
\Omega_0 = \d (p_A^\alpha v^A_\alpha-L)\wedge\d^mx-
\d p_A^\alpha\wedge\d y^A\wedge\d^{m-1}x_\alpha,
\end{gather*}
$({\cal W}_0,\Omega_0)$ (equiv. $({\cal W}_r,\hat h^*\Omega_0)$)
is a pre-multisymplectic Hamiltonian system.

\subsection{Field equations}

A {\it Lagrange--Hamilton problem} consists in f\/inding sections
$\psi_0\in\Gamma(M,{\cal W}_0)$ such that
  \begin{equation}
  \psi_0^*\inn(Y_0)\Omega_0=0  ,\qquad
  \forall \, Y_0\in\vf({\cal W}_0)   .
    \label{psi0}
\end{equation}
Taking $Y_0\in\vf^{{\rm V}(\hat\rho_2^0)}({\cal W}_0)$
we get the {\it first constraint submanifold}
$\jmath_1\colon{\cal W}_1\hookrightarrow{\cal W}_0$,
\[
{\cal W}_1 = \{ (\bar y,{\bf p})\in{\cal W}_0 \; | \;
\inn(V_0)(\Omega_0)_{({\bar y},{\bf p})}=0,\
\mbox{for every $V_0\in{\rm V}(\hat\rho_2^0 )$} \} ,
\]
and sections solution to (\ref{psi0}) take values on it.
${\cal W}_1$ is def\/ined by
$p^\alpha_A=\derpar{L}{v^A_\alpha}$,
hence
\[
{\cal W}_1=\{ (\bar {y},\widetilde{\cal F}\Lag(\bar y))\in{\cal W}\, \mid \,
\bar y\in J^1\pi\}\ ,
\]
and ${\cal W}_1$ is dif\/feomorphic to $J^1\pi$.

\begin{theorem} \emph{(see diagram \eqref{diagra})}
If $\psi_0\colon M\to{\cal W}_0$ is a section fulfilling equation
\eqref{psi0}, then $\psi_0=(\psi_\Lag,\psi_{\cal
H})=(\psi_\Lag,\widetilde{\cal F}\Lag\circ\psi_\Lag)$, where
$\psi_\Lag=\rho_1^0\circ\psi_0$, and:
 \begin{enumerate}\itemsep=0pt
  \item[$1.$] $\psi_\Lag$ is
the canonical lift of the projected section
$\phi=\rho_E^0\circ\psi_0\colon M \to E$ $($that is, $\psi_\Lag$ is
a~holonomic section$)$.
 \item[$2.$] $\psi_\Lag=j^1\phi$ is a
solution to the Lagrangian problem, and
$\mu\circ\psi_{\cal H}=\mu\circ\widetilde{\cal
FL}\circ\psi_\Lag={\cal F}\Lag\circ j^1\phi$ is a solution to the
Hamiltonian problem.

Conversely, for every section $\phi\colon M\to E$ such that
$j^1\phi$ is a solution to the Lagrangian problem $($and hence ${\cal
FL}\circ j^1\phi$ is a solution to the Hamiltonian problem$)$ we have
that $\psi_0=(j^1\phi,\widetilde{{\cal F}\Lag}\circ
j^1\phi)$, is a solution to \eqref{psi0}.
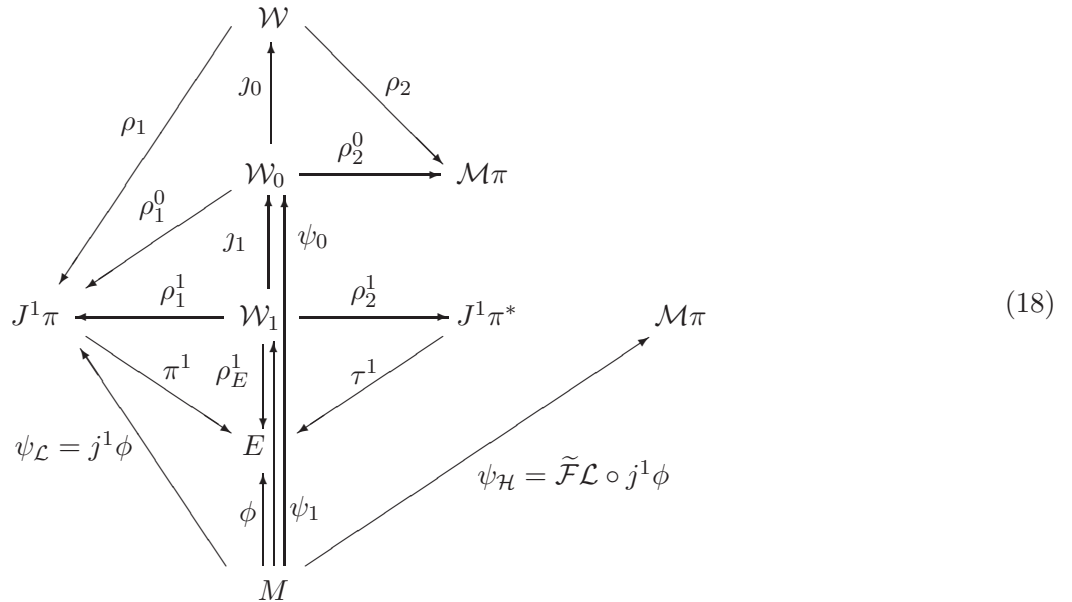
\begin{equation}
\begin{array}{cccc}
& \begin{picture}(135,100)(0,0)
\put(65,89){\mbox{${\cal W}$}}
\put(13,50){\mbox{$\rho_1$}}
\put(55,89){\vector(-2,-3){65}}
\put(58,65){\mbox{$\jmath_0$}}
\put(70,45){\vector(0,1){38}}
\put(83,89){\vector(1,-1){52}}
\put(113,65){\mbox{$\rho_2$}}
\put(60,30){\mbox{${\cal W}_0$}}
\put(140,30){\mbox{${\cal M}\pi$}}
\put(20,18){\mbox{$\rho^0_1$}}
\put(55,27){\vector(-3,-2){55}}
\put(52,6){\mbox{$\jmath_1$}}
\put(69,-10){\vector(0,1){35}}
\put(81,33){\vector(1,0){53}}
\put(95,40){\mbox{$\rho^0_2$}}
\end{picture} & &
\\
J^1\pi &
\begin{picture}(135,20)(0,0)
\put(28,10){\mbox{$\rho^1_1$}}
\put(52,3){\vector(-1,0){56}}
\put(58,0){\mbox{${\cal W}_1$}}
\put(100,10){\mbox{$\rho^1_2$}}
\put(81,3){\vector(1,0){56}}
\end{picture}
&
\begin{picture}(80,20)(0,0)
\put(-5,0){\mbox{$J^1\pi^*$}}
\put(70,0){\mbox{${\cal M}\pi$}}
\end{picture}
\\ &
\begin{picture}(135,100)(0,0)
\put(29,84){\mbox{$\pi^1$}}
\put(49,84){\mbox{$\rho_E^1$}}
\put(100,82){\mbox{$\tau^1$}}
\put(-27,55){\mbox{$\psi_\Lag=j^1\phi$}}
 \put(148,44){\mbox{$\psi_{\cal H}=\widetilde{\cal F}\Lag\circ j^1\phi$}}
\put(77,32){\mbox{$\psi_1$}}
\put(80,135){\mbox{$\psi_0$}}
 \put(58,30){\mbox{$\phi$}}
\put(59,55){\mbox{$E$}}
 \put(65,0){\mbox{$M$}}
\put(67,97){\vector(0,-1){32}}
\put(0,100){\vector(3,-2){55}}
 \put(135,100){\vector(-3,-2){55}}
\put(53,13){\vector(-2,3){55}}
 \put(83,13){\vector(3,2){130}}
\put(67,13){\vector(0,1){35}}
 \put(71,13){\vector(0,1){85}}
 \put(75,13){\vector(0,1){140}}
\end{picture} &
\begin{picture}(10,100)(0,0)
\end{picture} &
\end{array}
\label{diagra}
\end{equation}
\end{enumerate}
\label{mainteor1}
\end{theorem}

\begin{proof}
This proof is taken from  \cite{ELMMR-2004}. See also \cite{LMM-2002}.

1.~Taking $\big\{\derpar{}{p_A^\alpha}\big\}$
 as a local basis for the $\rho^0_1$-vertical vector f\/ields,
 and a section $\psi _0$, we have
 \begin{gather*}
 \inn\left(\derpar{}{p_A^{\alpha}}\right)\Omega_0  =
 v^A_{\alpha}\d^mx-\d y^A\wedge\d^{m-1}x_\alpha
 \quad \Longrightarrow \\
0=\psi _0^*\left[\inn\left(\derpar{}{p_A^\alpha}\right)\Omega_0\right]=
\left(v^A_\alpha(x)-\derpar{y^A}{x^\alpha}\right)\d^mx  ,
\end{gather*}
and thus the holonomy condition appears naturally within the
unif\/ied formalism,
 and it is not necessary to impose it by hand to $\psi _0$.
Thus we have that
$\psi_0=\big(x^\alpha,y^A,\derpar{y^A}{x^\alpha},\derpar{L}{v^A_\alpha}\big)$,
since $\psi_0$ takes values in ${\cal W}_1$, and hence it is of
the form $\psi_0=(j^1\phi,\widetilde{\cal FL}\circ j^1\phi)$, for
$\phi=(x^\alpha,y^A)=\rho_E^0\circ\psi_0$.

2.~Since sections
$\psi_0\colon M\to{\cal W}_0$ solution to (\ref{psi0}) take values
in ${\cal W}_1$, we can identify them with sections $\psi_1\colon
M\to{\cal W}_1$. These sections $\psi_1$ verify, in particular,
that $\psi_1^*\inn(Y_1)\Omega_1=0$ holds for every
$Y_1\in\vf({\cal W}_1)$. Obviously $\psi_0=\jmath_1\circ\psi_1$.
Moreover, as ${\cal W}_1$ is the graph of $\widetilde{\cal FL}$,
denoting by $\rho_1^1=\rho_1^0\circ\jmath_1\colon{\cal W}_1 \to
J^1\pi$ the dif\/feomorphism which identif\/ies ${\cal W}_1$ with
$J^1\pi$, if we def\/ine $\Omega_1=\jmath_1^*\Omega_0$, we have that
$\Omega_1=\rho_1^{1*}\Omega_\Lag$. In fact; as
$(\rho_1^1)^{-1}(\bar y)=(\bar y,\widetilde{{\cal F}\Lag}(\bar
y))$, for every $\bar y\in J^1\pi$, then
$(\rho_0^2\circ\jmath_1\circ(\rho_1^1)^{-1})(\bar y)=
\widetilde{{\cal F}\Lag}(\bar y)\in{\cal M}\pi$, and hence
\[
\Omega_\Lag=
\big(\rho_0^2\circ\jmath_1\circ(\rho_1^1)^{-1}\big)^*\Omega=
\big[\big(\big(\rho_1^1\big)^{-1}\big)^*\circ\jmath_1^*\circ\rho_0^{2*}\big]\Omega=
\big[\big(\big(\rho_1^1\big)^{-1}\big)^*\circ\jmath_1^*\big]\Omega_0=
\big(\big(\rho_1^1\big)^{-1}\big)^*\Omega_1   .
\]
Now, let $X\in\vf(J^1\pi)$. We have
\begin{gather}
(j^1\phi)^*\inn(X)\Omega_\Lag =
(\rho_1^0\circ\psi_0)^*\inn(X)\Omega_\Lag=
(\rho_1^0\circ\jmath_1\circ\psi_1)^*\inn(X)\Omega_\Lag
\nonumber \\
\phantom{(j^1\phi)^*\inn(X)\Omega_\Lag}{} =
(\rho_1^1\circ\psi_1)^*\inn(X)\Omega_\Lag=
\psi_1^*\inn((\rho_1^1)_*^{-1}X)(\rho_1^{1*}\Omega_\Lag)=
\psi_1^*\inn(Y_1)\Omega_1
\nonumber \\ \phantom{(j^1\phi)^*\inn(X)\Omega_\Lag}{}=
\psi_1^*\inn(Y_1)(\jmath_1^*\Omega_0)=
(\psi_1^*\circ\jmath_1^*)\inn(Y_0)\Omega_0=
\psi_0^*\inn(Y_0)\Omega_0   ,
\label{chain}
\end{gather}
where $Y_0\in\vf({\cal W}_0)$ is such that $Y_0=\jmath_{1*}Y_1$.
But as $\psi_0^*\inn(Y_0)\Omega_0=0$, for every $Y_0\in\vf({\cal W}_0)$,
then we conclude that $(j^1\phi)^*\inn(X)\Omega_\Lag=0$,
for every $X\in\vf(J^1\pi)$.

Conversely, let $j^1\phi\colon M\to J^1\pi$ such that
$(j^1\phi)^*\inn(X)\Omega_\Lag=0$, for every $X\in\vf(J^1\pi)$,
and def\/ine $\psi_0\colon M\to{\cal W}_0$ as
$\psi_0:=(j^1\phi,\widetilde{{\cal F}\Lag}\circ j^1\phi)$
(observe that $\psi_0$ takes its values in ${\cal W}_1$).
Taking into account that, on the points of ${\cal W}_1$,
every $Y_0\in\vf({\cal W}_0)$ splits into
$Y_0=Y_0^1+Y_0^2$, with $Y_0^1\in\vf({\cal W}_0)$ tangent to ${\cal W}_1$, and
$Y_0^2\in\vf^{{\rm V}(\rho_1^0)}({\cal W}_0)$, we have that
\[
\psi_0^*\inn(Y_0)\Omega_0=\psi_0^*\inn(Y_0^1)\Omega_0+\psi_0^*\inn(Y_0^2)\Omega_0=0  ,
\]
because for $Y_0^1$, the same reasoning as in (\ref{chain}) leads to
\[
\psi_0^*\inn(Y_0^1)\Omega_0=(j^1\phi)^*\inn(X_0^1)\Omega_\Lag=0  ,
\]
(where $X_0^1=(\rho_1^1)^{-1}_*Y_0^1$)
and, as $j^1\phi$ is a holonomic section for $Y_0^2$, following also the same reasoning as in (\ref{chain}),
a local calculus gives
\[
\psi_0^*\inn(Y_0^2)\Omega_0=
(j^1\phi)^*\left[\left(f_A^\alpha(x)\left( v_\alpha^A-\derpar{y^A}{x^\alpha}\right)\right)
\d^mx\right]=0   .
\]

The result for the sections ${\cal FL}\circ j^1\phi$
is a direct consequence of the {\it equivalence Theorem}~\ref{equiv1} between the Lagrangian and Hamiltonian formalisms.
\end{proof}

Thus, equation (\ref{psi0}) gives  equations of three dif\/ferent classes:
 \begin{enumerate}\itemsep=0pt
\itemsep 0pt plus 1pt
  \item
 Algebraic equations,
determining ${\cal W}_1\hookrightarrow{\cal W}_0$, where the
sections solution take their va\-lues. These are the
{\it primary Hamiltonian constraints}, and generate, by
$\hat\rho_2^0$ projection, the primary constraints of the
Hamiltonian formalism for singular Lagrangians.
\item
Dif\/ferential equations, forcing the sections
solution $\psi_0$ to be holonomic.
 \item
  The Euler--Lagrange equations.
 \end{enumerate}

Field equations in the unif\/ied formalism can also be stated in terms
of multivector f\/ields and connections in ${\cal W}_0$. In fact,
 the problem of f\/inding sections solution to (\ref{psi0})
  can be formulated equivalently as follows:
  f\/inding a distribution $D_0$ of $\Tan ({\cal W}_0)$ such that
  it is integrable (that is, {\it involutive\/}),
  $m$-dimensional, $\rho_M^0$-transverse, and
  the integral manifolds of $D_0$ are the sections solution
  to the above equations. (Note that we do not ask them to be lifting
  of $\pi$-sections; that is, the holonomic condition).
  This is equivalent to stating that the sections solution to
  this problem are the integral sections of one of the following
  equivalent elements:
\begin{itemize}\itemsep=0pt
\item
A class of integrable and $\rho_M^0$-transverse $m$-multivector f\/ields
  $\{ X_0\}\subset\vf^m({\cal W}_0)$ satisfying that
  \[
    \inn (X_0)\Omega_0=0   ,   \qquad
  \mbox{\rm for every $X_0\in\{ X_0\}$}   .
 \]
 \item
An integrable connection $\nabla_0$
in $\rho_M^0\colon{\cal W}_0\to M$ such that
\[
  \inn(\nabla_0)\Omega_0=(m-1)\Omega_0  .
 \]
 \end{itemize}
Locally decomposable and $\rho_M^0$-transverse multivector f\/ields
and orientable connections which are solutions of these equations
are called {\it Lagrange--Hamiltonian multivector fields} and
{\it jet fields} for $({\cal W}_0,\Omega_0)$.
Euler--Lagrange and Hamilton--De Donder--Weyl multivector f\/ields
can be recovered from these Lagrange--Hamiltonian multivector f\/ields
(see \cite{ELMMR-2004}).

\section{Example}
\protect\label{examp}

As an example of application of these formalisms we consider
a classical system which has been taken from \cite{ELMMR-2004}: minimal surfaces (in $\Real^3$).
Other examples of application of the multisymplectic formalism are explained in detail in
\cite{GMS-97,GIMMSY-mm,Sd-95} as well as in many other references
(see, for instance,
\cite{CCI-91,EM-92,EMR-96,EMR-98,EMR-99b,EMR-00,
LMM-96,LMM-2002} and quoted references).

\subsection{Geometric elements. Lagrangian and Hamiltonian formalisms}

The problem consists in looking for mappings
$\varphi\colon U\subset\Real^2\to\Real$ such that their graphs have minimal area
as sets of $\Real^3$, and satisfy certain boundary conditions.

For this model, we have that $M=\Real^2$, $E=\Real^2\times\Real$, and
\begin{gather*}
J^1\pi=\pi^*\Tan^*\Real^2\otimes\Real=\pi^*\Tan^*M=\pi^*\Tan^*\Real^2
  , \qquad
{\cal M}\pi=\pi^*(\Tan M\times_ME)
 , \\
J^1\pi^*=\pi^*\Tan M=\pi^*\Tan\Real^2   .
\end{gather*}
The coordinates in $J^1\pi$, $J^1\pi^*$ and ${\cal M}\pi$ are denoted
$(x^1,x^2,y,v_1,v_2)$, $(x^1,x^2,y,p^1,p^2)$, and $(x^1,x^2,y,p^1,p^2,p)$
respectively. If $\omega=\d x^1\wedge\d x^2$, the Lagrangian density is
\[
\Lag=\big[1+(v_1)^2+(v_2)^2\big]^{1/2}\d x^1\wedge\d x^2\equiv \lag\d x^1\wedge\d x^2   ,
\]
and the Poincar\'e--Cartan forms are
\begin{gather*}
\Theta_\Lag =  \frac{v_1}{\lag}\d y\wedge\d x^2-\frac{v_2}{\lag}\d y\wedge\d x^1+
\lag\left( 1-\left(\frac{v_1}{\lag}\right)^2-\left(\frac{v_2}{\lag}\right)^2\right)
\d x^1\wedge\d x^2,
\\
\Omega_\Lag =  -\d\left(\frac{v_1}{\lag}\right)\wedge\d y\wedge\d x^2+
\d\left(\frac{v_2}{\lag}\right)\wedge\d y\wedge\d x^1\\
\phantom{\Omega_\Lag =}{} -
\d\left[ \lag\left( 1-\left(\frac{v_1}{\lag}\right)^2-\left(\frac{v_2}{\lag}\right)^2\right)\right]
\wedge\d x^1\wedge\d x^2   .
\end{gather*}
The Euler--Lagrange equation of the problem are
\begin{gather}
0=\left(\derpar{p^2}{x^2}+\derpar{p^1}{x^1}\right)\d x^1\wedge\d x^2=
\left[\derpar{}{x^1}\left(\frac{v_1}{\lag}\right)+\derpar{}{x^2}\left(\frac{v_2}{\lag}\right)\right]
\d x^1\wedge\d x^2
\nonumber
\\
\phantom{0}{}=\frac{1}{\lag^3}
\Bigg[\left( 1+\left(\derpar{y}{x^1}\right)^2\right)\frac{\partial^2y}{\partial x^2\partial x^2}
+\left( 1+\left(\derpar{y}{x^2}\right)^2\right)\frac{\partial^2y}{\partial x^1\partial x^1}\nonumber\\
\phantom{0=}{}-
2\derpar{y}{x^1}\derpar{y}{x^2}\frac{\partial^2y}{\partial x^1\partial x^2}\Bigg]
\d x^1\wedge\d x^2  ,
\label{eulag}
\end{gather}
and the associated Euler--Lagrange $m$-vector f\/ields and connections
 which are the solutions to the Lagrangian problem are
\begin{gather*}
   X_\Lag  =  f
  \left(\derpar{}{x^1}+v_1\derpar{}{y}+
  \derpar{v_1}{x^1}\derpar{}{v_1}+\derpar{v_2}{x^1}\derpar{}{v_2}\right)\wedge
  \left(\derpar{}{x^2}+v_2\derpar{}{y}+
  \derpar{v_1}{x^2}\derpar{}{v_1}+\derpar{v_2}{x^2}\derpar{}{v_2}\right),
   \\
\nabla_\Lag  =  \d x^1\otimes
\left(\derpar{}{x^1}+v_1\derpar{}{y}+
\derpar{v_1}{x^1}\derpar{}{v_1}+\derpar{v_2}{x^1}\derpar{}{v_2}\right)
\\ \phantom{\nabla_\Lag  =}{}+
\d x^2\otimes
\left(\derpar{}{x^2}+v_2\derpar{}{y}+
\derpar{v_1}{x^2}\derpar{}{v_1}+\derpar{v_2}{x^2}\derpar{}{v_2}\right)   .
\end{gather*}

The Legendre maps are given by
\begin{gather*}
{\cal F}\Lag(x^1,x^2,y,v_1,v_2) = \left(x^1,x^2,y,\frac{v_1}{\lag},\frac{v_2}{\lag}\right),
\\
\widetilde{{\cal F}\Lag}(x^1,x^2,y,v_1,v_2) =
\left( x^1,x^2,y,\frac{v_1}{\lag},\frac{v_2}{\lag},\lag-\frac{(v_1) ^2}{\lag}-\frac{(v_2)^2}{\lag}\right)  ,
\end{gather*}
and then $\Lag$ is hyperregular. The Hamiltonian function is
${\rm h}=-[1-(p^1)^2-(p^2)^2]^{1/2}$, and so the Hamilton--Cartan forms are
\begin{gather*}
\Theta_h =  p^1\d y\wedge\d x^2-p^2\d y\wedge\d x^1-{\rm h}\d x^1\wedge\d x^2,
\\
\Omega_h =  -\d p^1\wedge\d y\wedge\d x^2+\d p^2\wedge\d y\wedge\d x^1+
\d {\rm h}\wedge\d x^1\wedge\d x^2   .
\end{gather*}
The Hamilton--De Donder--Weyl equations of the problem are
\begin{gather}
\derpar{y}{x^1}=-\frac{p^1}{{\rm h}}  , \qquad
\derpar{y}{x^2}=-\frac{p^2}{{\rm h}}, \qquad
\derpar{p^1}{x^1}=-\derpar{p^2}{x^2}   ,
\label{hamdeweyl}
\end{gather}
and the corresponding Hamilton--De Donder--Weyl $m$-vector f\/ields and connections
 which are the solutions to the Hamiltonian problem are
\begin{gather*}
   X_h  =  f
  \left(\derpar{}{x^1}-\frac{p^1}{{\rm h}}\derpar{}{y}+
  \derpar{p^1}{x^1}\derpar{}{p^1}+
  \derpar{p^2}{x^1}\derpar{}{p^2}\right)\wedge
  \left(\derpar{}{x^2}-\frac{p^2}{{\rm h}}\derpar{}{y}+
  \derpar{p^1}{x^2}\derpar{}{p^1}+
  \derpar{p^2}{x^2}\derpar{}{p^2}\right),
   \\
\nabla_h  =  \d x^1\otimes
\left(\derpar{}{x^1}-\frac{p^1}{{\rm h}}\derpar{}{y}+
\derpar{p^1}{x^1}\derpar{}{p^1}+
\derpar{p^2}{x^1}\derpar{}{p^2}\right)
\\ \phantom{\nabla_h  =}{} +
\d x^2\otimes
\left(\derpar{}{x^2}-\frac{p^2}{{\rm h}}\derpar{}{y}+
  \derpar{p^1}{x^2}\derpar{}{p^1}+
  \derpar{p^2}{x^2}\derpar{}{p^2}\right)   .
\end{gather*}

\subsection{Unif\/ied formalism}

For the unif\/ied formalism we have
\[
{\cal W}=\pi^*\Tan^*M\times_E\pi^*(\Tan M\times_ME)
  ,\qquad
{\cal W}_r=\pi^*\Tan^*M\times_E\pi^*\Tan M=\pi^*(\Tan^*M\times_M\Tan M)   .
\]
If $w=(x^1,x^2,y,v_1,v_2,p^1,p^2,p)\in{\cal W}$, the coupling form is
$\hat{\cal C}=(p^1v_1+p^2v_2+p)\d x^1\wedge\d x^2$;
therefore
\[
{\cal W}_0=\big\{ (x^1,x^2,y,v_1,v_2,p^1,p^2,p)\in{\cal W}\, \mid\,
[1+(v_1)^2+(v_2)^2]^{1/2}-p^1v_1-p^2v_2-p=0\big\}   ,
\]
and we have the forms
\begin{gather*}
\Theta_0 =
\big(\big[1+(v_1)^2+(v_2)^2\big]^{1/2}-p^1v_1-p^2v_2\big)\d x^1\wedge\d x^2-
p^2\d y\wedge\d x_1+p^1\d y\wedge\d x_2,
\\
\Omega_0 =
-\d \big(\big[1+(v_1)^2+(v_2)^2\big]^{1/2}-p^1v_1-p^2v_2\big)\wedge\d x^1\wedge\d x^2\\
\phantom{\Omega_0 =}{} +
\d p^2\wedge\d y\wedge\d x_1-\d p^1\wedge\d y\wedge\d x_2  .
\end{gather*}
Taking f\/irst $\hat\rho_2^0$-vertical vector f\/ields $\derpar{}{v_\alpha}$
we obtain
\[
0=\inn\left(\derpar{}{v_\alpha}\right)\Omega_0=
\left(p^\alpha-\frac{v_\alpha}{\lag}\right)\d x^1\wedge\d x^2   ,
\]
which determines the submanifold
${\cal W}_1={\rm graph}\,\widetilde{{\cal F}\Lag}$ (dif\/feomorphic to $J^1\pi$),
and reproduces the expression of the Legendre map.
Now, taking $\rho^0_1$-vertical vector f\/ields $\derpar{}{p^\alpha}$,
the contraction $\inn\big(\derpar{}{p^\alpha}\big)\Omega_0$ gives,
for $\alpha=1,2$,
$ v_1\d x^1\wedge\d x^2-\d y\wedge\d x^2$ and
$v_2\d x^1\wedge\d x^2+\d y\wedge\d x^1$ respectively,
so that, for a section
$\psi_0=(x^1,x^2,y(x^1,x^2),v_1(x^1,x^2),v_2(x^1,x^2),p^1(x^1,x^2),p^2(x^1,x^2))$
taking values in ${\cal W}_1$, we have that the condition
$\psi _0^*\big[\inn\big(\derpar{}{p^\alpha}\big)\Omega_0\big]=0$
leads to
\[
\left(v_1-\derpar{y}{x^1}\right)\d x^1\wedge\d x^2=0   , \qquad
\left(v_2-\derpar{y}{x^2}\right)\d x^1\wedge\d x^2=0   ,
\]
which are the holonomy condition. Finally, taking the vector f\/ield $\derpar{}{y}$ we have
\[
\inn\left(\derpar{}{y}\right)\Omega_0=-\d p^2\wedge\d x^1+\d p^1\wedge\d x^2
\]
and, for a section $\psi_0$ fulf\/illing the former conditions,
the equation $  0=\psi_0^*\big[\inn\big(\derpar{}{y}\big)\Omega_0\big]$ leads to
 the Euler--Lagrange equations (\ref{eulag}).
Now, bearing in mind the expressions of ${\rm h}$ and the Legendre map,
from the Euler--Lagrange equations we get
 the Hamilton--De Donder--Weyl equations~(\ref{hamdeweyl}).

The $m$-vector f\/ields and connections which are the
solutions to the problem in the unif\/ied formalism are
 \begin{gather*}
   X_0  =  f
  \left(\derpar{}{x^1}+v_1\derpar{}{y}+
  \derpar{v_1}{x^1}\derpar{}{v_1}+\derpar{v_2}{x^1}\derpar{}{v_2}+
  \derpar{p^1}{x^1}\derpar{}{p^1}+\derpar{p^2}{x^1}\derpar{}{p^2}\right)
\\ \phantom{X_0  =}{} \wedge
  \left(\derpar{}{x^2}+v_2\derpar{}{y}+
  \derpar{v_1}{x^2}\derpar{}{v_1}+\derpar{v_2}{x^2}\derpar{}{v_2}+
  \derpar{p^1}{x^2}\derpar{}{p^1}+\derpar{p^2}{x^2}\derpar{}{p^2}\right),
   \\
\nabla_0  =  \d x^1\otimes
\left(\derpar{}{x^1}+v_1\derpar{}{y}+
\derpar{v_1}{x^1}\derpar{}{v_1}+\derpar{v_2}{x^1}\derpar{}{v_2}+
\derpar{p^1}{x^1}\derpar{}{p^1}+\derpar{p^2}{x^1}\derpar{}{p^2}\right)
\\ \phantom{\nabla_0  =}{} +
\d x^2\otimes\left(\derpar{}{x^2}+v_2\derpar{}{y}+
\derpar{v_1}{x^2}\derpar{}{v_1}+\derpar{v_2}{x^2}\derpar{}{v_2}+
\derpar{p^1}{x^2}\derpar{}{p^1}+\derpar{p^2}{x^2}\derpar{}{p^2}\right)   ,
\end{gather*}
($f$ being a non-vanishing function)
where the coef\/f\/icients $\derpar{v_\alpha}{x^\nu}=
\frac{\partial^2 y}{\partial x^\nu\partial x^\alpha}$
are related by the Euler--Lagrange equations, and the coef\/f\/icients
$\derpar{p^\alpha}{x^\nu}$ are related by the Hamilton--De Donder--Weyl equations
(the third one).
From these expressions we recover the
 Euler--Lagrange $m$-vector f\/ields and connections
 which are the solutions to the Lagrangian problem, and
 the Hamilton--De Donder--Weyl $m$-vector f\/ields and connections
 which are the solutions to the Hamiltonian problem
 obtained in the above paragraph.

\section{Discussion and outlook}
\protect\label{conclusion}

Multisymplectic geometry and its application to describe classical f\/ield theories
have been  f\/ields of increasing interest in the last years.
A lot of well-known results in the realm of symplectic geometry and symplectic mechanics have been
generalized also for the multisymplectic case, but
there are many other problems which remain open.
Next we review some of these results and problems, and their current status.

A fundamental result in symplectic geometry is the Darboux theorem.
The analogous result also holds in some particular cases of multisymplectic forms
(for instance, for volume forms).
Nevertheless, in the general case, a multisymplectic manifold
does not admit a system of Darboux coordinates for the multisymplectic form.
In fact this is a problem arising from linear algebra: the classif\/ication
of skew-symmetric tensors of degree greater than two is still an open problem.
The kind of multisymplectic manifolds
admitting Darboux coordinates has been identif\/ied \cite{LMS-2003},
and they are those being locally multisymplectomorphic to bundles of forms
(see also \cite{FG-2007} for another approach to this problem).

Another interesting subject concerns to the def\/inition of {\it Poisson brackets}
in multisymplectic manifolds. This is a relevant point, for instance,
for the further quantization of classical f\/ield theories.
This problem has been studied in the realm of polysymplectic manifolds
\cite{Ka-95,Ka-97b} and for the multisymplectic case some recent contributions are
\cite{FPR-2003,FPR-2003b,FPR-2005}.
However, the problem is not completely solved satisfactorily,
and the research on this topic is still open.

In the same way, approaches for generalizing symplectic integrators
to this geometric framework (i.e., the so-called {\it multisymplectic integrators\/})
have been studied in recent years,
and numerical methods have been
developed for solving the f\/ield equations, which are based on
the use of these multisymplectic integrators \cite{MGS-99,MS-99}.
Research on this topic is in progress.

Another f\/ield of increasing interest in the last years is the study of
systems in classical f\/ield theories with nonholonomic constraints.
This is a meeting topic between honholonomic mechanics and classical f\/ield theories.
The construction of the Lagrangian and Hamiltonian formalism, as well
as other problems such as the study of symmetries and reduction have been analyzed
for the $k$-symplectic formulation \cite{LMSV-2009}
and for the multisymplectic models in several works
\cite{BLMS,Van-1,Van-2,Van-3,Van-4}.

Further developments have not been achieved.
For instance,  the generalization of the
Mars\-den--Weinstein reduction theorem~\cite{MW-74} to the multisymplectic framework.
Concerning reduction theo\-ry in general, only partial results
about reduction by foliations are currently being studied~\cite{Ib-2000}.
The corresponding reduction theorem has been stated and proved
for the $k$-symplectic formulation~\cite{fam},
but the theory of reduction of multisymplectic Lagrangian and Hamiltonian systems under the action of
groups of symmetries is still under research, and only partial results have been achieved
\cite{CGR-2001,CM-2003,CRS,MW}.

The problem of quantization of classical f\/ield theories is another relevant topic to be deve\-lo\-ped.
There are several works due to Kanatchikov devoted to geometric (pre)quantization
of polysymplectic f\/ield theories
\cite{Ka-q1,Ka-q2,Ka-q3,Ka-q4,Ka-q4b,Ka-q5,Ka-q6},
some attempts for the $k$-symplectic case~\cite{B-2008,PCM-2001}, and other dif\/ferent approaches for
the quantization of f\/ields, in general (see, for instance,~\cite{Ba-2004,Sd-94}).
Nevertheless, the study of the geometric structures and obstructions to
perform the geometric quantization program for
covariant multisymplectic f\/ield theories is open to further research.

As a f\/inal remark, many of the subjects that we have presented in
this work  have been studied also for higher-order f\/ield theories
(see, for instance, \cite{AA-78,AA-80a,FF-86,FF-89,GM-83,KS-2000,Kr-84,Sa-89,SC-90}).
One of the problems of the f\/irst multisymplectic models for these theories
was that the def\/inition of the corresponding multisymplectic structure (the Poincar\'e--Cartan form)
was ambiguous. This trouble have been solved recently
\cite{CLMV-2009}. But, in general, the problem of stating complete
and satisfactory geometrical models for the Lagrangian and Hamiltonian formalisms
of these kinds of theories, as well as other related topics
(symmetries, constraint algorithms for the singular cases, quantization, \ldots)
 are under development.

One can expect to see more work on all these subjects in the future.

\appendix

\section{Appendix}

\subsection{Multisymplectic manifolds}
\protect\label{mf}

 \begin{definition}
 Let ${\cal M}$ be a dif\/ferentiable manifold, and
  $\Omega\in\df^k({\cal M})$ ($1<k\leq\dim {\cal M}$).

  $\Omega$ is a {\it multisymplectic form},
and then $({\cal M},\Omega)$ is a {\it multisymplectic manifold}, if
  \begin{enumerate}\itemsep=0pt
  \item
  $\Omega\in Z^k({\cal M})$ (it is closed).
  \item
  $\Omega$ is $1$-nondegenerate; that is,
for every $p\in{\cal M}$ and $X_p\in\Tan_p{\cal M}$,
  $\inn(X_p)\Omega_p=0$ $\Leftrightarrow$ $X_p=0$.
  \end{enumerate}
 If $\Omega$ is closed and $1$-degenerate then it is a {\it
 pre-multisymplectic form}, and $({\cal M},\Omega)$
 is a {\it pre-multisymplectic manifold}.
\end{definition}

Multisymplectic manifolds of degree $k=2$
are the usual symplectic manifolds, and manifolds
with a distinguished volume form are
multisymplectic manifolds of degree its dimension.
Other examples of multisymplectic manifolds are provided by
compact semisimple Lie groups equipped with the canonical
cohomology 3-class, symplectic 6-dimensional Calabi--Yau manifolds
with the canonical 3-class, etc.
There are no multisymplectic manifolds of degrees $1$ or $\dim\,{\cal M}-1$
because $\ker\Omega$ is nonvanishing in both cases.

 Another very important kind of
multisymplectic manifold is the {\it multicotangent bundle}
of a manifold $Q$, $\Lambda^k(\Tan^*Q)$, that is,
the bundle of $k$-forms in $Q$. This bundle
is endowed with a~canonical $k$-form
$\Theta\in\df^k(\Lambda^k(\Tan^*Q)$, and then
$\Omega:=-\d\Theta\in\df^{k+1}(\Lambda^k(\Tan^*Q)$
is a $1$-nondegenerate form. Then the couple
$(\Lambda^k(\Tan^*Q),\Omega)$ is a multisymplectic manifold.

A local classif\/ication of multisymplectic forms can be done only
for particular cases \cite{LMS-2003,FG-2007}.

\subsection{Multivector f\/ields}
\protect\label{mvf}

See \cite{EMR-98} for details.
Let ${\cal M}$ be a $n$-dimensional dif\/ferentiable manifold.
Sections of $\Lambda^m(\Tan {\cal M})$ are called
$m$-{\it multivector fields} in ${\cal M}$
(they are the contravariant skew-symmetric tensors of order~$m$ in~${\cal M}$).
We denote by $\vf^m ({\cal M})$ the set of
$m$-multivector f\/ields in~${\cal M}$. Then,
${\cal X}\in\vf^m({\cal M})$ is {\it locally decomposable} if,
for every $p\in {\cal M}$, there is an open neighbourhood
$U_p\subset {\cal M}$
and $X_1,\ldots ,X_m\in\vf (U_p)$ such that
${\cal X}\vert_{U_p}=X_1\wedge\ldots\wedge X_m$.

A non-vanishing ${\cal X}\in\vf^m({\cal M})$ and
a $m$-dimensional distribution ${\cal D}\subset\Tan {\cal M}$
are {\it locally associated} if there exists a connected open set
$U\subseteq {\cal M}$ such that ${\cal X}\vert_U$
is a section of $\Lambda^m{\cal D}\vert_U$.
If ${\cal X},{\cal X}'\in\vf^m({\cal M})$ are non-vanishing multivector f\/ields
locally associated with the same distribution ${\cal D}$,
on the same connected open set $U$, then there exists a
non-vanishing function $f\in\Cinfty (U)$ such that
${\cal X}'\vert_{U}=f{\cal X}$. This fact def\/ines an equivalence relation in the
set of non-vanishing $m$-multivector f\/ields in ${\cal M}$,
whose equivalence classes
will be denoted by $\{ {\cal X}\}_U$. Then
there is a~one-to-one correspondence between the $m$-dimensional
orientable distributions ${\cal D}$ in $\Tan {\cal M}$ and the
equivalence classes $\{ {\cal X}\}_{\cal M}$
 of non-vanishing, locally decomposable
$m$-multivector f\/ields in~${\cal M}$.

A non-vanishing, locally decomposable
multivector f\/ield ${\cal X}\in\vf^m({\cal M})$ is said to be {\it integrable}
(resp.\ {\it involutive}) if
 its associated distribution is integrable (resp.\ involutive).
If ${\cal X}\in\vf^m({\cal M})$ is integrable (resp.\ involutive),
then so is every other in its equivalence class $\{ {\cal X}\}$,
and all of them have the same integral manifolds.
Moreover, {\it Frobenius theorem} allows us to say that
a non-vanishing and locally decomposable multivector f\/ield is integrable
 if, and only if, it is involutive.

If $\pi\colon {\cal M}\to M$ is a f\/iber bundle,
we are interested in the case where the integral manifolds of
integrable multivector f\/ields in ${\cal M}$ are sections of $\pi$.
Thus, ${\cal X}\in\vf^m({\cal M})$ is said to be {\it $\pi$-transverse}
if, at every point $y\in {\cal M}$,
$(\inn ({\cal X})(\pi^*\beta))_y\not= 0$, for every $\beta\in\df^m(M)$
with $\omega (\pi(y))\not= 0$.
Then, if ${\cal X}\in\vf^m({\cal M})$ is integrable,
it is $\pi$-transverse if, and only if,
its integral manifolds are local sections of $\pi\colon {\cal M}\to M$.
Finally, it is clear that  classes of locally decomposable and
$\pi$-transverse multivector f\/ields $\{{\cal X}\} \subseteq \vf^m ({\cal M})$
are in one-to-one correspondence with orientable Ehresmann connection
forms $\nabla$ in $\pi\colon{\cal M}\to M$. This correspondence is
characterized by the fact that the horizontal subbundle associated
with $\nabla$ is the distribution associated with $\{ {\cal X}\}$.
In this correspondence,
classes of integrable locally decomposable and $\pi$-transverse
$m$ multivector f\/ields correspond to f\/lat orientable Ehresmann
connections.

\subsection*{Acknowledgements}

I acknowledge the f\/inancial support of
\emph{Ministerio de Educaci\'on y Ciencia},
projects
MTM\,2005--04947, MTM\,2008--00689/MTM and MTM\,2008--03606--E/MTM.
I wish to thank to Professors Miguel~C.~Mu\~noz-Lecanda
and Xavier Gr\`acia for their comments,
and to the referees, whose suggestions have allowed me to improve this work.
Finally, thanks also to Mr.~Jef\/f Palmer for his
assistance in preparing the English version of the manuscript.

\pdfbookmark[1]{References}{ref}

\LastPageEnding

\end{document}